\documentclass[aps,showpacs,twocolumn,floatfix,letterpaper,citeautoscript,superscriptaddress
,10pt]{revtex4-1}

\usepackage[intlimits,sumlimits]{amsmath}
\usepackage{amsfonts,amssymb}
\usepackage{bm}
\usepackage{graphicx}
\usepackage{hyperref}
\usepackage[vcentermath]{youngtab}
\usepackage{array}
\usepackage{multirow}

\newcommand{\be}{\begin{equation}}
\newcommand{\ee}{\end{equation}}
\newcommand{\ba}{\begin{equation}\begin{split}}
\newcommand{\ea}{\end{split}\end{equation}}

\newcommand{\bs}{\boldsymbol}

\catcode`,\active

\catcode`\,12

\hypersetup{%
pdftitle={Theory of Transport Phenomena in Coherent Quantum Hall Bilayers},%
pdfauthor={Inti Sodemann, Hua Chen and Allan H. MacDonald},%
pdfpagemode={UseNone},%
pdfstartview={FitH},%
breaklinks=true}

\begin{document}
\title{Theory of Transport Phenomena in Coherent Quantum Hall Bilayers}

\author{I. Sodemann}
\affiliation{Department of Physics, Massachusetts Institute of Technology, Cambridge, Massachusetts 02139}
\author{Hua Chen}
\affiliation{Department of Physics, University of Texas at Austin, Austin, Texas 78712}
\author{A. H. MacDonald}
\affiliation{Department of Physics, University of Texas at Austin, Austin, Texas 78712}

\date{\today}

\begin{abstract}
We argue that all anomalous transport properties of coherent quantum quantum Hall bilayers can be understood in terms of a mean-field transport theory in which the condensate phase is nearly uniform across the sample, and the strength of condensate coupling to interlayer tunneling
processes is substantially reduced compared to the predictions of disorder-free microscopic mean-field theory. 
These ingredients provide a natural explanation for recently established I-V characteristics which feature a critical current 
above which the tunneling resistance abruptly increases and a non-local interaction between interlayer tunneling at the 
inner and outer edges of Corbino rings. We propose a microscopic picture in which disorder is the main agent responsible for the 
reduction of the effective interlayer tunneling strength.    
\end{abstract}
\maketitle


\section{Introduction} 
The discovery a number of years ago by Spielman {\it et al.}~\cite{Spielman2000} of a huge
enhancement of the zero-bias tunneling conductance between two nearby two-dimensional-electron layers 
provided compelling evidence for an interesting 
ordered state in which interlayer phase coherence is established spontaneously~\cite{Eisenstein2004,Eisenstein2014}. 
This tunneling anomaly has so far been observed only in the quantum Hall regime 
and only near total Landau level filling factor $\nu_{T}=1$.
The ordered state can be viewed as a condensate formed from electrons in the lowest Landau level of one layer and holes in the lowest Landau level of the other layer.
Other striking transport anomalies are characteristic the same 
ordered state, including quantized Hall drag~\cite{Kellogg2002}, 
vanishing Hall and longitudinal resistances~\cite{Kellogg2004,Tutuc2004,Wiersma2004} when Hall bar bilayers are contacted so that opposite layers 
carry currents in opposite directions, and excitonic Andreev scattering~\cite{Finck2011} and nearly perfect 
Coulomb drag in Corbino geometry bilayers~\cite{Nandi2012}.  

A bilayer state with spontaneous interlayer phase coherence is equivalent to a state with a condensate of 
spatially indirect excitons.  Prior to Spielman {\it et al.}'s work
the appearance of this unusual type of broken symmetry in quantum Hall bilayers
had been predicted~\cite{Fertig1989} on the basis of microscopic considerations 
particular to the quantum Hall regime, hinted at experimentally~\cite{Eisenstein1992,Murphy1994},
and studied theoretically~\cite{MacDonald1990,Brey1990,Wen1992,Wen1993,Narikiyo1993,Ezawa1993,Yang1994,Ezawa1994,Moon1995,Yang1996}.
The most spectacular properties of these states were however
not revealed until transport measurements  were undertaken in which the two 
layers were independently contacted.
The first of these~\cite{Spielman2000}
revealed the tunneling anomaly and stimulated a 
considerable body of additional theoretical work~\cite{Stern2001,Balents2001,Fogler2001,Joglekar2001,Yang2001,Demler2001,Stern2002,Fertig2003,
Sheng2003,Fertig2005,Rossi2005,Park2006,Su2008,Eastham2009,Fil2009,Dolcini2010,Eastham2010,Su2010,Hyart2011,Pesin2011,Hyart2013,Hama2013}. 

\begin{figure}
\includegraphics[width=3.0in]{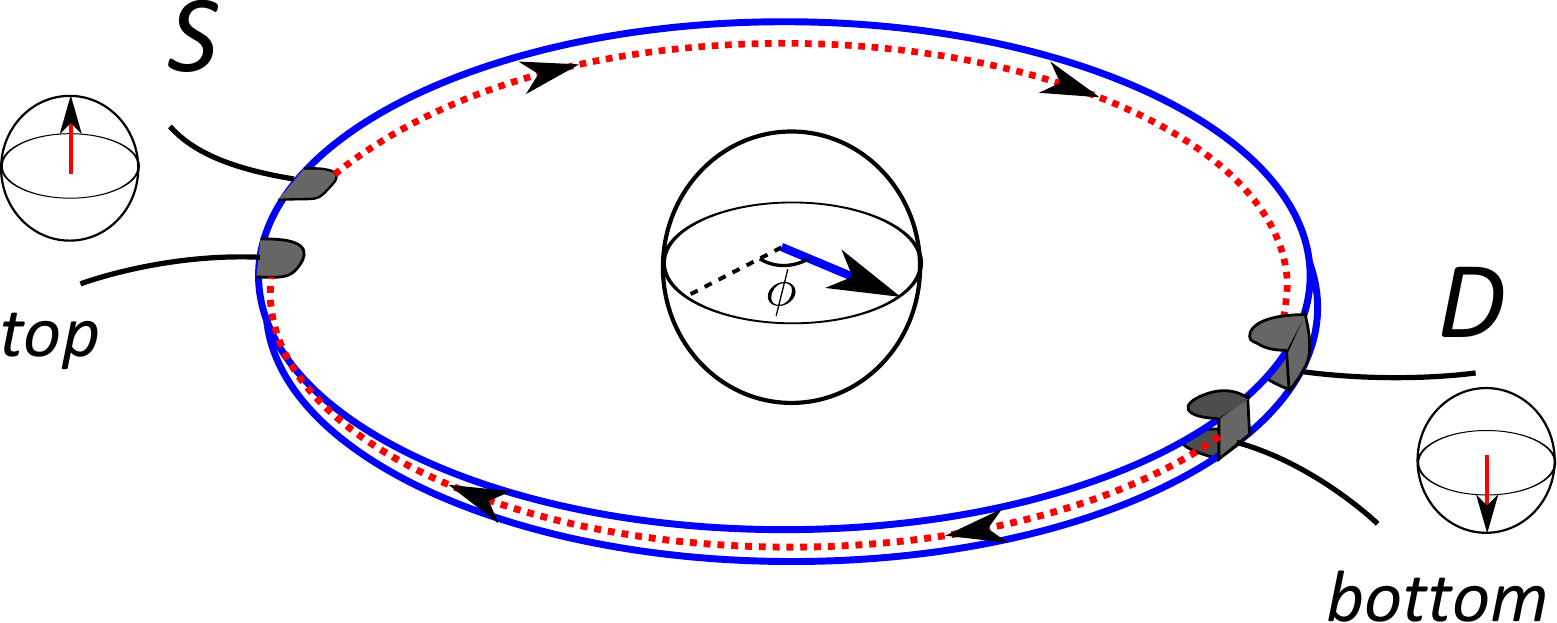}
\caption{Schematic illustration of the anomalous tunneling phenomena in coherent quantum Hall bilayers. 
A high chemical potential source is connected to the top layer and a low chemical potential drain is connected to the bottom layer. 
The layer pseudospin of transport electrons rapidly rotates from the layer-polarized source and drain orientations to the
layer-coherent condensate orientations over small regions close to the contacts depicted by gray shading. 
Quasiparticle states in the source and drain electrodes are layer-localized and have pseudo spins 
which lie close to the north pole (for top layer electrodes) or to the south pole (for bottom layer states). 
The edge and bulk quasiparticle states in the condensate have coherent layer-delocalized wave functions and 
layer pseudo spins which lie close to the XY  plane with a phase.  Current experiments are 
consistent with an interlayer phase $\phi$ that is nearly uniform throughout the sample.
Well formed quantum Hall plateaus demonstrate that transport electrons flow along sample edges.
Corbino geometry samples have independent transport at inner and outer edges.  
}
\label{Fig1}
\end{figure}

The relatively small value of the zero-bias tunneling conductance peak observed in early experiments,
and its relatively weak dependence on in-plane magnetic field~\cite{Spielman2001,Spielman2004}, 
motivated theoretical researchers~\cite{Stern2001,Balents2001,Fogler2001,Fertig2003,
Sheng2003,Fertig2005,Eastham2009,Fil2009,Eastham2010,Hyart2011,Hyart2013} to develop 
descriptions that started from the assumption that the interlayer coherent state
lacked long-range order due to a combination of quantum and thermal order parameter fluctuations, and disorder. 
However, more recent experiments~\cite{Spielman2004,Tiemann2008,Tiemann2009,Yoon2010,Finck2011,Nandi2012,Huang2012,Nandi2013,Zhang2014},
often using four terminal measurement configurations to separate contact and tunneling resistances, have made it 
increasingly clear that the zero-bias tunnel conductance is at least as large as $\sim 250 e^2/h$ and possibility much larger.
These experiments also provided strong evidence for the presence of a 
critical inter-layer current which becomes sharp in the low temperature limit and 
is reminiscent of the critical current of a Josephson junction.  When the critical current is exceeded
the tunneling resistance abruptly increases.
Most tellingly, experiments in Corbino geometry samples~\cite{Huang2012,Nandi2013} 
have revealed 
a non-local interaction between tunneling characteristics 
at the sample's inner and outer edges.  In particular, these experiments demonstrated that 
the critical current is determined by the sum~\cite{Su2008} of the separate interlayer currents at the 
two independent edges.   As we will explain, these results shed new light on the nature of the bilayer coherent state and directly demonstrate long-range coherence of the inter-layer phase. 
 
We argue below that all transport anomalies which accompany quantum Hall bilayer exciton condensation (see Fig.~\ref{Fig1}),
including the tunneling anomaly, can be described by combining a mean-field theory
of electron-electron interactions with independent quasiparticle 
transport theory~\cite{Wen1993,Su2008,Pesin2011} and accounting for
the influence of excitonic Andreev scattering on the interlayer phase~\cite{Su2008}. 
We also propose new experiments which explore the time dependence of the condensate current-driven-dynamics and 
can be used both to further test our picture of bilayer exciton condensate transport anomalies
and to quantify the order parameter energy functional it employs.  
 

Our paper is organized as follows.  In Section II we review the central ideas needed to understand the transport anomalies of bilayer exciton condensates.  In 
Section III we describe transport in the macrospin limit in which the interlayer phase
is constant across the sample, demonstrating that theoretical predictions in this limit
achieve agreement with experiment for a variety of different measurement configurations.
The spatial properties of the condensate depend on whether the length scale $\lambda$, defined by 
comparing the pseudospin stiffness parameter in the energy functional $\rho$ with the tunneling parameter $\Delta$ which 
characterizes energetic preference for a particular interlayer phase value, is larger or smaller than the system size. 
The macrospin limit applies when $\lambda$ is larger than the system size. 
In Section IV we consider the case in which $\lambda$ is smaller than the system size.
We demonstrate that the predictions of theory for this limit  are in qualitative disagreement with recent experiments
which have found the condensate state to be controlled by the sum of the currents
at the inner and outer edges of Corbino-geometry samples. 
In Section V we explore the time dependence of the condensate dynamics, a regime of 
phenomena where experiments are still lacking.  
Comparison of theory and experiment for dynamical properties 
could be used to further deepen our understanding of bilayer exciton condensates, and 
in particular to test the theoretical framework we propose here.  
When microscopic mean-field theory is applied to disorder-free samples, the 
values predicted for $\Delta$ and $\rho$ would not place experimental samples in the 
macrospin limit. In Section VI we discuss a picture for the microscopic origin of the dramatic reduction
in $\Delta$ required to resolve this discrepancy.  Its most important conclusion is that the phase appearing in Sections II-V should  
be understood as a coarse grained phase averaged over distances that are long compared to the disorder correlation length. Finally, we present our conclusions and summarize our findings in Section VII.

\section{Transport Theory} 

The phenomena we address have close analogies with spin-transfer torques and related effects in magnets and with tunneling in superconducting Josephson junctions. An order parameter for the pseudospin {\it which layer} degree-of-freedom can be defined as follows: 

\be
{\bf M}\equiv \frac{1}{2 A}\sum_{k,\alpha\beta} {\bs \sigma}_{\alpha\beta}\langle c_{k\alpha}^\dagger c_{k\beta}\rangle,
\ee

\noindent where $c^{\dagger},c$ are electron creation and annihilation operators,
$k$ is an intra-Landau-level orbital index, $\alpha,\beta$ run over top ($t$) and bottom ($b$) layer labels, and $A$ is the area of the system.  
Following a standard procedure, an action describing the low energy dynamics can be constructed in terms of a local order parameter ${\bf M}(\vec{r},t)$ whose magnitude remains constant while its orientation changes in space and time~\cite{Moon1995}. The corresponding Lagrangian density is:
\be
\mathcal{L}=\hbar \dot {\phi}(\vec{r},t) M_z(\vec{r},t)-\mathcal{E}[\phi(\vec{r},t),M_z(\vec{r},t)],
\ee

\noindent where $\phi$ is the azymuthal angle of the local order parameter and $M_z$ its projection along the z-axis, which is half of the electron density difference between the layers.  In the zero temperature mean-field theory of the disorder-free $\nu_{T}=1$ bilayer,
all electrons share a common pseudospin state
yielding a ground state with the maximum pseudospin polarization possible in the lowest Landau level Hilbert space: 
$|{\bf M}|= n/2$, where $n$ is the electron density. 
More generally, one expects the order parameter to be reduced compared to this value as we discuss at length in Section~\ref{Sec:disc}.

The capacitive energy is assumed large enough to keep the order parameter in-plane: $M_z\ll|{\bf M}|$. The energy per unit area
can then be written as: 

\be
\label{eq:3} 
\mathcal{E}[\phi,M_z]= - \Delta \cos \phi+\rho  |\nabla \phi|^2+\frac{e^2}{2 c} M_z^2.
\ee

\noindent  The first term arises from the single particle tunneling of electrons and it is controlled by the splitting, $\Delta_{SAS}$, between ($S$) and antisymmetric ($AS$) double well states: $\Delta=\Delta_{\rm SAS} |{\bf M}| $. The second term captures 
the energy cost of spatial variation in interlayer phase characterized by $\rho$, the stiffness constant.
The third term is the capacitive energy which is responsible for keeping the order parameter in the XY plane.
Here $c$ is the capacitance per unit area.  

The non-dissipative dynamics of the phase follows from the Euler-Lagrange equations associated with $\mathcal{L}$. Dissipation arises when incoherent tunneling processes are accounted for~\cite{Joglekar2001} and the connection to external reservoirs via electrical leads is included. 
The dynamics including these terms is described by:

\begin{equation}\begin{split}
\hbar \dot{{M_z}}=  -\Delta \sin\phi +2 \rho \nabla^2 \phi - \frac{g_{qp}}{A}  \, \hbar \dot{\phi}  + \sum_{k} \frac{\hbar  I_k}{2} l_k  \delta (\vec{r}-\vec{R}_k)
\end{split}.
\label{eq:open} 
\end{equation}

\noindent The first two terms in the right of the equation are non-dissipative and account respectively for coherent collective tunneling 
between layers and the divergence of the exciton supercurrent. The third term is a dissipative and 
accounts for incoherent tunneling between layers with a time-dependent interlayer phase.
For simplicity we assume here that it is Ohmic allowing it to be characterized by 
an associated conductance $g_{qp}$ measured in units of $e^2/\hbar$. 
This term can be understood by noting that $\hbar \dot{\phi}$ is equivalent to 
a chemical potential difference between the 
layers~\footnote{This is a direct consequence of the fact that $\phi$ and $M_z$ are canonically conjugated variables~\cite{Wen1993}:
the Euler-Lagrange equation for $M_z$ implies $\hbar\dot{\phi}=\delta \mathcal{E}/\delta M_z$, and the chemical potential difference between 
the layers is the energy cost of moving one electron from the bottom to the top layer, for which $\int d^2 r \ \delta M_z=1$.}.  It is the 
layer pseudospin analog of Gilbert damping in a ferromagnet~\cite{Gilbert2004}, and of the shunt resistance in the RCSJ 
model of a Josephson junction~\cite{Tinkham2012}. 
We have omitted an additional dissipative term proportional to $\nabla^2 \dot{\phi}$ that would arise 
when the bulk of the quantum Hall liquid is not perfectly insulating~\cite{Fogler2001}.  

The last term in Eq.~\eqref{eq:open} accounts for injection and removal of quasiparticles through layer-polarized 
contacts.  As in the familiar case of  superconducting Andreev scattering at NS interfaces, quasiparticle scattering 
off the order parameter violates particle number conservation (independent particle number in each layer 
in the present case) and is accompanied by a reaction effect which creates or annihilates excitons~\cite{Su2008}.
We assume the contacts are connected to the edge of the sample at positions $R_k$.
In Eq.~\eqref{eq:open}, $I_k$ is the electron number current injected into the sample at contact $k$, and $l_k$ takes values $\pm 1$ for top and bottom layer contacts respectively.   Assuming that quasiparticle interlayer coherence is established close to source and drain electrodes and  
integrating Eq.~\eqref{eq:open} over an small region $\Omega_k$ surrounding the contact~\cite{Su2008,Su2010}, we 
see that single-layer contacts effectively provide a Neumann type boundary condition for the normal derivative of the phase: 
\be\label{neumann}
l_k\frac{\hbar I_k}{2} = -\oint_{\partial \Omega_k} d s \ \hat{n}\cdot  (2 \rho \nabla\phi).
\ee 
Eq.~\eqref{neumann} expresses the idea that the layer polarized quasiparticle current is converted into an excitonic supercurrent over a short microscopic distance. The conversion process which entails the creation or annihilation of interlayer excitons close to current sources and drains is analogous to Andreev scattering in superconductors and to spin-transfer torques in ferromagnets~\cite{Su2008,Su2010}. 

So far we have discussed the equations which capture how transport currents influence the bilayer order  
parameter.  The converse effect in which bilayer order influences transport, which is further detailed 
in the following sections, is even stronger and proceeds in the first place by 
establishing coherence between layers.  
In the quantum Hall regime the charge current flows from source to drain along
chiral edge channels~\cite{Prange1987}.  
The source-to-drain conductance depends on contact details but is ideally $\sim e^2/h$~\cite{Pesin2011}. 
In a bilayer exciton condensate the edge states,
like bulk states, are interlayer-coherent and have nearly equal magnitude amplitudes in the two layers.
When the interlayer phase is time independent, voltage probes always 
measure the same voltage whether contacted to one layer or the other.  
When the interlayer phase is time-dependent but spatially constant, 
$\hbar \dot{\phi}/e$, contributes to interlayer voltages.  This last effect is 
closely analogous~\cite{Chen2014} to spin-pumping~\cite{Tserkovnyak2002,Tserkovnyak2005} in ferromagnets as we explain further below.   
Figure~\ref{Fig1} summarizes transport in the coherent state.

\section{Macrospin limit}
When the system is able to reach a time-independent steady state, the phase throughout the sample will satisfy a sine-Gordon equation:

\be\label{SineGordon}
\lambda^2 \nabla^2 \phi -\sin\phi=0
\ee

\noindent which must be solved together with the Neumann boundary conditions in Eq.~\eqref{neumann}. 
The length scale $\lambda=\sqrt{2\rho/\Delta}$ sets the typical scale over which the phase changes. 
In the limit in which the typical dimension of the sample is much smaller than $\lambda$,
the phase of the condensate remains uniform throughout the sample.
We refer to this as the {\it macrospin limit} in analogy with magnetism. 
In the macrospin limit the criterion for the existence of time-independent solutions to Eq.~\eqref{eq:open} reduces to:

\be\label{Icrit}
\left|\sum_k l_k \frac{\hbar I_k}{2}\right|\leq A \Delta, 
\ee

\noindent where $A$ is the area of the sample. $\sum_k l_k \hbar I_k/2$ can be interpreted as the net rate
at which layer pseudospin is transferred to the condensate by quasiparticle transport currents.  
Eq.~\ref{Icrit} predicts that the critical current is proportional to sample area, 
as established experimentally in Ref.~\onlinecite{Finck2008}, and that 
it is the {\it total} layer polarized current injected or extracted out of the sample over all current contacts
which controls the stability of time-independent solutions.  This conclusion applies 
even if the sample is multiply connected, as it is the case in the Corbino geometry, and the currents are being
injected simultaneously through otherwise disconnected edges. It is this recent experimental finding in Corbino samples~\cite{Huang2012,Nandi2013} which convincingly indicates that current experimental systems are effectively in the macrospin limit.  

In the remainder of this section we assume the macrospin limit and demonstrate that 
the theory we have outlined is able to reproduce the shape of two- and four-terminal I-V characteristics.
Assuming the system to have a sufficiently strong capacitive energy to suppress the $\dot{M_z}$ term  
in Eq.~\eqref{eq:open}, one obtains the following equation for the macrospin dynamics:

\begin{equation} 
\label{eq:macrospin} 
\frac{A \Delta}{\hbar}  \sin\phi   +  g_{qp} \, \dot{\phi}  - \sum_k l_k I_k/2  = 0.
\end{equation} 

\noindent In the two-terminal tunneling configuration where an electron number current $I$ is driven from a source 
terminal connected to the top layer to a drain terminal along the same edge that is connected to the bottom layer, we have:
$\sum_k l_k I_k/2=I$. Denoting the chemical potential difference between source and drain by $\mu$, one has:

\be\label{Iprecess}
\hbar I = g (\mu-\hbar \dot{\phi}),
\ee

\noindent where $g$ is the two-terminal conductance in $e^2/\hbar$ units. 
The shift of the source-to-drain bias by $-\hbar \dot{\phi}$ captures the influence of the order parameter 
dynamics on the charge quasiparticle current flowing through the system edges.    
The $-\hbar \dot{\phi}$  shift in the interlayer bias can be understood by performing a unitary transformation
to a rotating frame that transforms the quasiparticle Hamiltonian into a time-independent one~\cite{Pesin2011}. 
This effect is the analog of the spin-pumping in metal spintronics~\cite{Tserkovnyak2002,Tserkovnyak2005}, and gives rise to an effective enhancement of the damping factor for the condensate phase: $g_{qp}\rightarrow g_{qp}+g$ ~\cite{Wen1993}.

The phase of the order parameter is time independent for $I<I_c$, with $\hbar I_c=A \Delta$.  Over this range of current, 
the current-voltage characteristic is $\hbar I=g \mu$. 
When the bias exceeds a critical value $\mu>\mu_c\equiv\hbar I_c/g$, the phase starts to precess. 
By integrating Eq.~\eqref{eq:macrospin} with respect to time we find that the average precession rate is: 

\be\label{phidot_av}
\hbar \langle \dot{\phi} \rangle =\frac{g}{g_{qp} + g} \sqrt{\mu^2-\mu_c^2},
\ee

\noindent and that the two-terminal I-V characteristic is given at biases exceeding $\mu_c$ by

\be
\hbar I = g \Big( \mu - \frac{g}{g_{qp}+g} \, \sqrt{\mu^2-\mu_c^2} \, \Big).  
\ee

\begin{figure}
\includegraphics[width=3in]{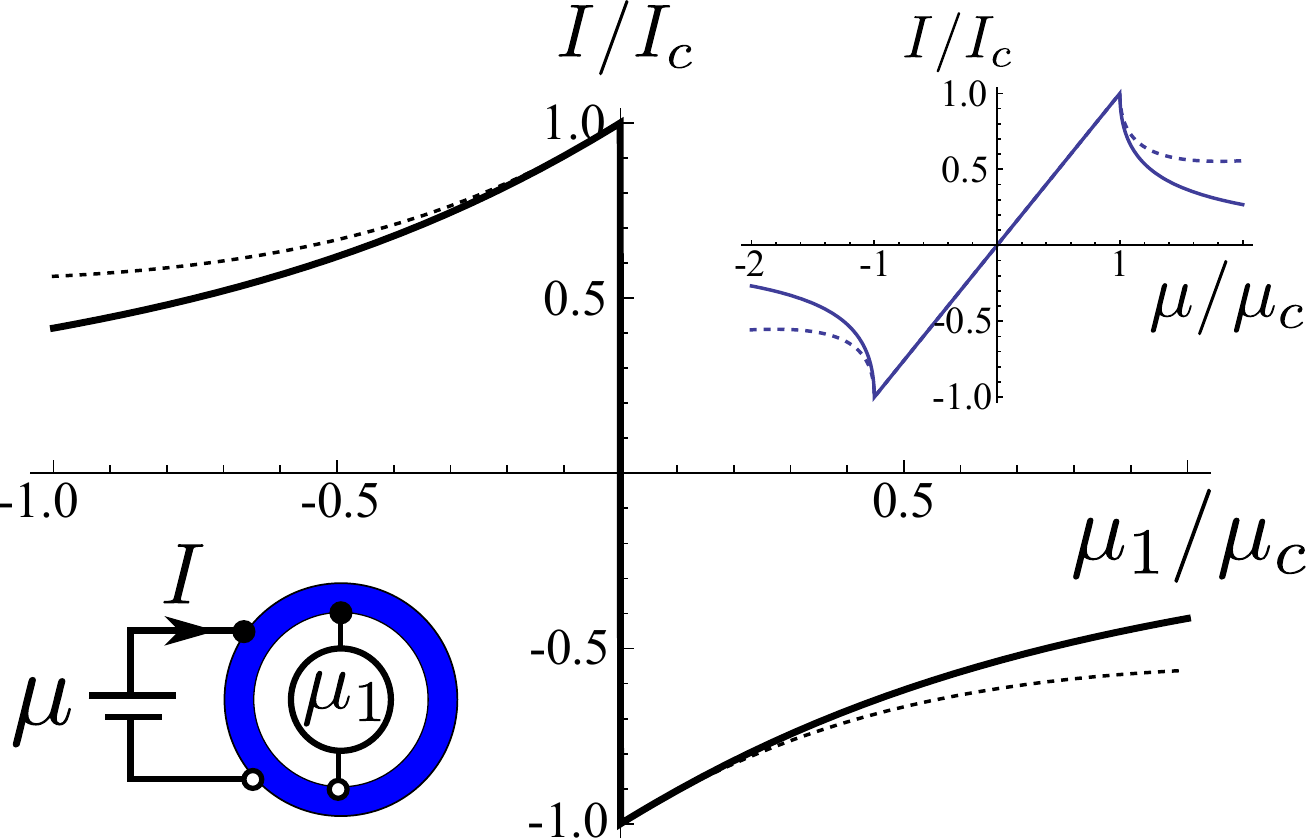}
\caption{(color online) Tunneling I-V characteristics in the macrospin limit. The lower inset provides a schematic
illustration of the measurement geometry. The filled black and empty white dots represent top and bottom layer contacts respectively.  
$\mu/e$ (plotted in the inset) is the source to drain bias voltage, and 
$\mu_1/e=\hbar \langle \dot{\phi} \rangle/e$ is the interlayer voltage measured 
at the inner edge when the tunneling current flows at the outer edge. 
The solid lines are for the limit $g_{qp} \rightarrow0$ and the dashed lines are for $g_{qp}=0.2 g$.
}
\label{IVs}
\end{figure}

As discussed previously $\hbar \langle \dot{\phi} \rangle$ is the chemical potential difference between layers
measured by four-terminal voltage probes.  If the transport current flows at the outer 
edge, it can be directly measured in a Corbino geometry by placing voltage 
probes at the inner edge of the device.  The resulting I-V characteristics, illustrated in Fig.~\ref{IVs}, display cusp-like features 
near a critical bias as observed experimentally~\cite{Tiemann2008,Tiemann2009,
Yoon2010,Nandi2013}~\footnote{The system I-V curve has a region of negative differential resistivity at voltages immediately above the
cusp which leads to circuit instabilities that make measurements extremely close to the maximum in the
I-V curve difficult.  Ref.~\onlinecite{Nandi2013} addressed this experimental challenge and provided
a thorough characterization of the behavior of the I-V curve immediately above the critical current.}.  
These cusp like features are absent~\cite{Nandi2013} in the theory of Ref.~\onlinecite{Hyart2011}. 
The values of $\Delta$, $g$ and $g_{qp}$ appropriate to a particular sample can be extracted simply by measuring the critical current, 
the critical voltage, and the interlayer voltage in the dynamic regime.  All experimental data of which we are aware can be 
accurately described in terms of these three parameters, although we expect small deviations in the dynamic 
regime because of the established non-ohmic character of incoherent interlayer tunneling in the quantum Hall regime.    

\begin{figure}
\includegraphics[width=3in]{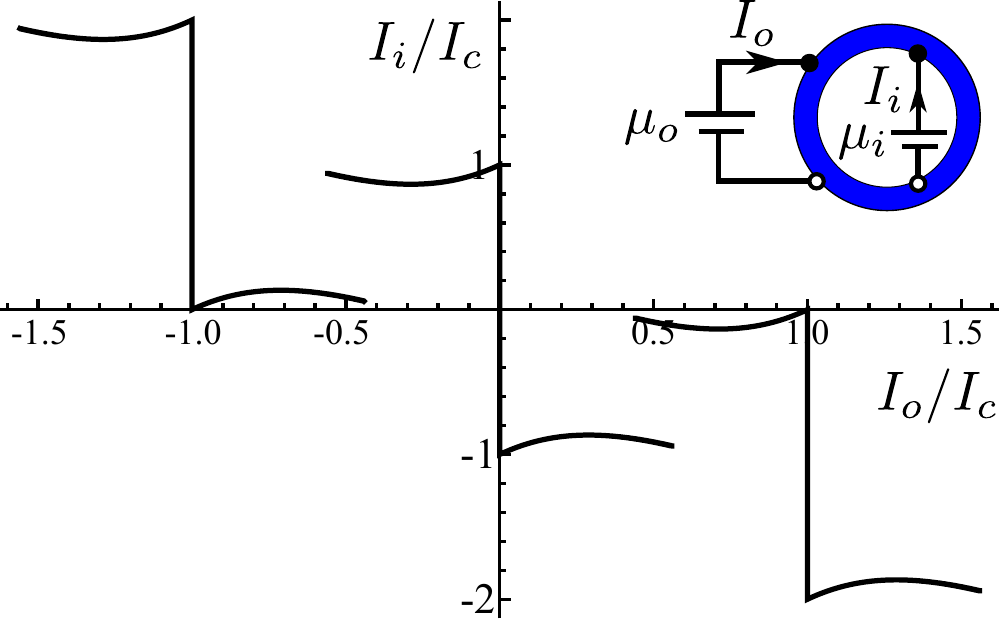}
\caption{Non-local coupling between tunneling at the inner and outer edges of a Corbino geometry sample in the macrospin limit. 
The three curves correspond are for bias sweeps at the outer edge $\mu_o$ at three different fixed biases in the inner edge: $\mu_i=\{-\mu_c,0,\mu_c\}$. 
In the static regime $I_{i} = g \mu_{i}$ is independent of $\mu_{o}$ and $I_{o}=g \mu_{o}$.  In the dynamic regime $I_{i}$ 
decreases with increases in $\mu_{o}$ because $\langle \dot{\phi} \rangle$ increases.  
For this illustration we have taken $g_{qp}\rightarrow 0$, and $g_i=g_o=g$.  The inset schematically 
illustrates the measurement geometry.  The filled black and empty white dots represent top and bottom layer contacts respectively. }
\label{I1_I2}
\end{figure}

Now we consider the experimental situation in which two sources inject independent tunneling currents $I_i$ and $I_o$ into the inner and outer edges of the Corbino device simultaneously~\cite{Huang2012,Nandi2013}.  We count both currents as positive when electrons 
are injected into the top layer and collected from the bottom layer.  The rates of pseudospin transfer in Eq.~\eqref{eq:macrospin} add up: $\sum_k l_k I_k/2=I_i+I_o$. Therefore, the criterion for the existence of a time independent condensate phase is simply: $|I_i+I_o|<I_c$. 
Quasiparticle transport at inner and outer edges is characterized by $\hbar I_{i,o}=g_{i,o}(\mu_{i,o}-\hbar \dot{\phi})$, with $g_{i,o}$ and $\mu_{i,o}$ the  conductances (in $e^2/\hbar$ units) and the chemical potential biases driving the currents at each edge. 
Note that the effective bias reduction, $-\hbar \dot{\phi}$, is identical at inner and outer edges because 
condensate phase precession is spatially uniform in the macrospin limit. 
Figure~\ref{I1_I2} illustrates predictions of the macrospin theory for 
this experimental geometry.  These curves are in good agreement with experiment~\cite{Huang2012,Nandi2013} and 
in particular capture two essential properties of the I-V characteristics measured in this situation:
i) cusps in the I-V curves occur at a fixed value of $I_{i}+I_{o}=\mu_{i}/g_{i}+\mu_{o}/g_{i}$ and 
ii) conductance measured at the outer edge is larger in the dynamic regime in the 
present geometry than in the tunneling ($g_{i} \to 0$) limit.  The latter property 
can be understood by viewing the conducting link between the upper and lower layers at
the inner edge as a contribution to incoherent tunneling.  The former property 
directly establishes the applicability of the macrospin limit.     
This connection will be made more explicit in Section~\ref{finiterho} where we 
calculate critical lines in the $I_{i}-I_{o}$ plane for $\lambda$ smaller than sample sizes.
Although surprising for reasons explained below, the conclusion that current 
samples are in the macrospin limit is inescapable.  

The drag and series counterflow geometries identified in Ref.~\cite{Su2008} are special cases of the 
two bias configurations described above, and are also successfully described by our theory.
Consider first the nearly perfect counterflow Coulomb drag observed in the experiments of Ref.~\cite{Nandi2012}. 
In these experiments a current $I_1$ is driven through the top layer between the inner and outer edges of a corbino annulus (drive current), while a current $I_2$ is measured between the inner and outer edges of the bottom layer (drag current). The experiment found that $I_1$ and $I_2$ are nearly equal in magnitude and oppositely directed.  When the conductivity $\sigma_{xx}$ of the quantum Hall bulk is zero, the 
only quasiparticle transport path consists of tunneling from top to bottom at the outer edge by interacting with the 
condensate, followed by 
current flow through the conducting link between the inner and outer edges of the bottom layer, followed 
by tunneling from bottom layer to top layer by interacting once again with the condensate.  It follows that 
$I_1 = I_2$; the small difference observed experimentally is easily accounted for by imperfect insulating behavior of the quantum Hall bulk (i.e. by 
$\sigma_{xx}$ having a small non-zero value).  
In this configuration the total rate of layer-pseudospin injected into the bilayer vanishes, i.e. $\sum_k l_k I_k/2=0$. Therefore the condensate phase remains static and close to its equilibrium value: $\phi=0$ up to large currents. 
The equality of the drive and drag currents follows from the layer delocalized character of 
edge states in the presence of the condensate which produces an effective short-circuit between the drive and drag loops at the edges. 

\begin{figure}
\includegraphics[width=3in]{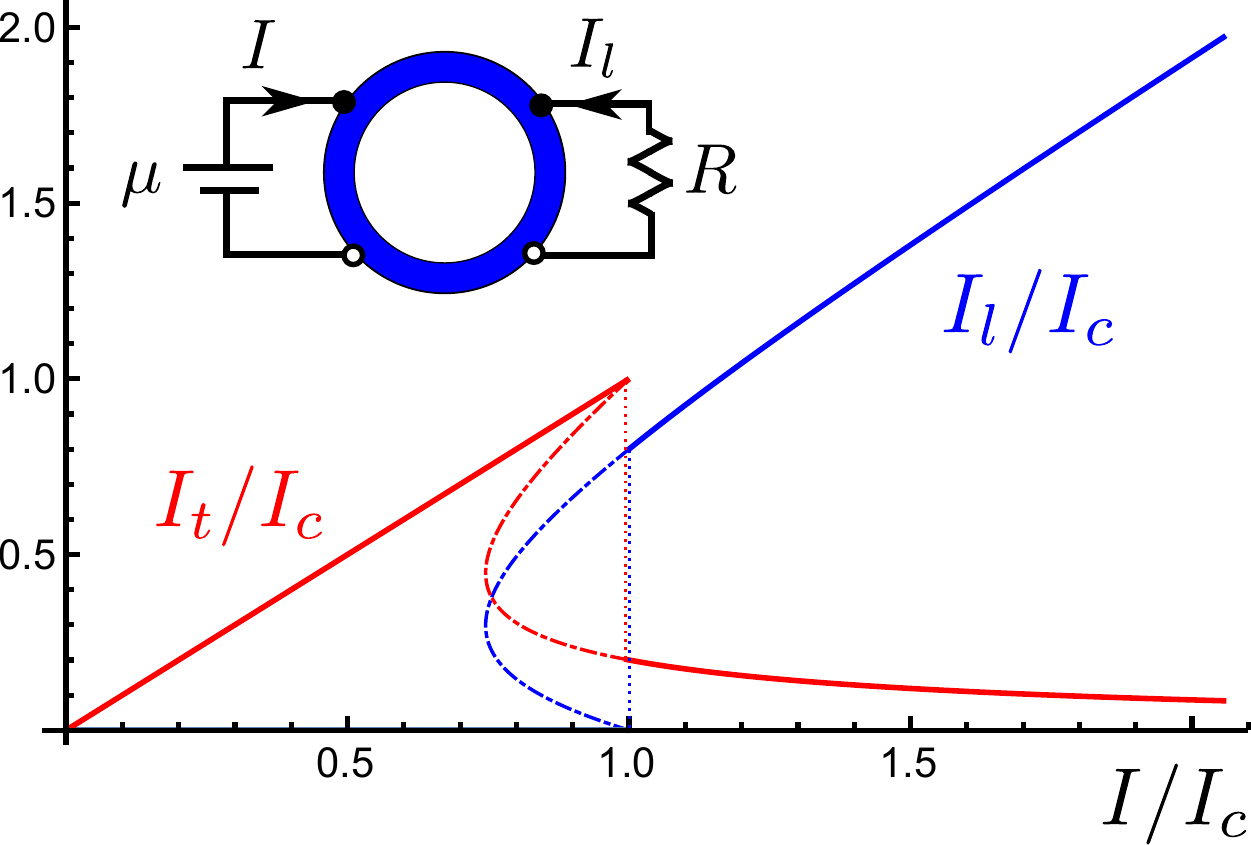}
\caption{(color online) Series counterflow measurement. The inset depicts the measurement configuration. The filled black dots and the empty white dots represent top and bottom layer contacts respectively. The tunneling current, defined as $I_t\equiv I -I_l$, is shown in red and the loop current, $I_l$, in blue. The dashed dotted lines indicate ranges over which the curves are multivalued. We have taken $g_{qp}\rightarrow 0$, and $g=g_l=1/R$ for 
this illustration.  In the low bias voltage regime $I_l=0$ whereas in the high bias voltage regime $I_l \approx I$.   The steady state established 
at high bias voltages is one in which $\hbar \dot{\phi}$ drives almost all the current through the loop resistor.    
}
\label{counter}
\end{figure}

In the series counterflow configuration investigated in Ref.~\cite{Yoon2010}, the sample was a Hall bar.  
A current $I$ is driven by a bias voltage $\mu/e$ between source and drain contacts to top and bottom layers at one
end of the Hall bar and a conducting link between top and bottom layers is provided at the 
other end of the Hall bar.  (See inset in Fig.~\ref{counter}). 
The source current and bias are related by $\hbar I=g (\mu-\hbar \dot{\phi})$, with $g_L$ the two-terminal conductance measured between source and drain contacts in the absence of the loop resistor. The chemical potential drop 
across the loop resistor, $\mu_l$ is related to the loop current, $I_l$, by Ohm's law: $\hbar I_l= R \mu_l$, with $R$ the loop resistance
in units of $\hbar/e^2$.  This chemical potential difference determines the tunneling current driven into 
the bilayer by the loop via: $\hbar I_l=-g_l (\mu_l-\hbar\dot{\phi})$~\footnote{The minus sign arises because the positive loop resistance sign convention has been redefined to be be positive when the number current is injected into the bottom layer to match the convention of Ref.~\cite{Yoon2010}.}, with $g_l$ the two-terminal conductance of the bilayer measured at the loop contacts (ideally $e^2/h$). 
This implies that the loop current directly probes of the condensate precession rate:

\be
\dot{\phi}=(R+g_l^{-1}) I_l.
\ee  
 
\noindent Combining these equations with the property that net layer pseudospin transfer is $\sum_k l_k I_k/2=I-I_l$, one obtains 
the behavior illustrated in Fig.~\ref{counter}, in agreement with experiment.  For currents below the critical current,
the finite loop resistor plays no role and the I-V characteristic is identical to that of a two-contact tunneling geometry.  
When the critical current is exceeded, a steady state is established in which $\hbar \dot{\phi}$ drives a quasiparticle current from 
bottom to top at the right end of the Hall bar which is completed by current flowing through the external loop.

Our theory also explains all other transport anomalies in a very simple manner.
In the Hall drag measurement~\cite{Kellogg2002} current flows between source and drain contacts connected 
along the same edge and to the 
same layer, the drive layer. Because both contacts are made to the same layer, there is no net pseudsopin
transfer torque and the order is always static.  All voltage measurements, including in particular Hall voltage measurements,
would measure a value that is independent of the contacting layer.  The Hall voltage in particular has the same value in the 
drag layer and the drive layer, in agreement with experiment.  The Hall voltages measured in 
series counterflow experiments~\cite{Tutuc2004,Wiersma2004} vanish because the net current 
carried along the Hall bar is zero.  In these experiments, voltage probes obtain non-zero values 
equal to $\hbar \langle \dot{\phi} \rangle/e$ in magnitude only when attached to opposite layers.

\section{Critical currents for finite condensate stiffness}\label{finiterho}

For systems with larger values of $\Delta$ or smaller values of $\rho$ the length scale $\lambda$ which determines the typical distance over which the condensate phase changes can become smaller than the system size and the macrospin picture will no longer
be appropriate. In this section we explore the regime in which the phase changes substantially across 
the sample.  One immediate conclusion from this study is that 
results for this limit are 
inconsistent with the measurements of Refs.~\cite{Huang2012,Nandi2013}. 
Specifically, we find that the boundary of the stability region for solutions with time independent condensate phase 
differs dramatically from the region $|I_i+I_o| = I_c$, predicted in the macrospin limit and 
found experimentally in Corbino device~\cite{Huang2012,Nandi2013} measurements.

\begin{figure}[!]
\includegraphics[scale=0.6]{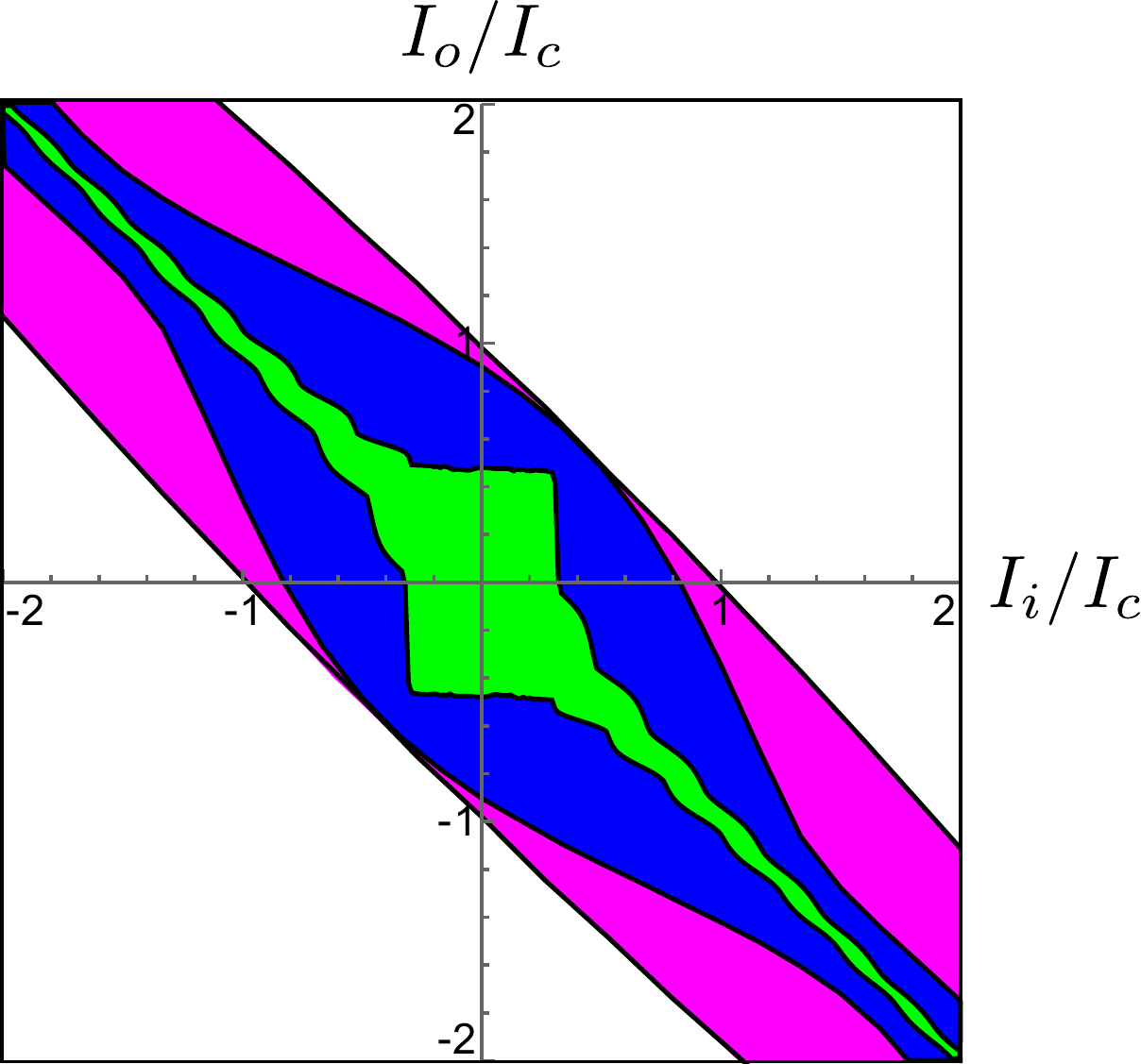}
\caption{\label{Ii-Io-corbino} (color online).
Area in the $I_{i}-I_{o}$ parameter space for which a static condensate steady state exists.  
The three different regions correspond to $\lambda=R_i$ (purple), $R_i/2$ (blue), $R_i/5$ (green), and the Corbino ring was chosen to have $R_o=2R_i$. 
For $\lambda=R_i$ the macrospin description is already accurate.}     
\end{figure}

When the length scale $\lambda$ is smaller than the system size, determining the critical currents requires detailed modeling of the sample geometry and 
in particular of the spatial arrangement of the contacts~\footnote{In this sense the very observation of the independence of the critical current on the contact position along the rim of the Corbino annulus is already 
an indication that $\lambda$ is longer than the system size~\cite{Nandi2013}.}. For simplicity we employ a toy model of current injection
by assuming that the tunneling current injection and exciton creation and annihilation occur uniformly along inner and outer Corbino 
edges with radii $R_i$ and $R_o$.  This geometry is easier to analyze because 
any dependence on azimuthal angle is ruled out by symmetry. In this case Eq.~\ref{SineGordon} reduces to:

\be\label{RadialSG}
\frac{\lambda^2}{r} \frac{d}{dr} \left(r\frac{d\phi}{dr}\right)=\sin \phi,
\ee

\noindent and the boundary conditions from Eq.~\ref{neumann} reduce to:

\be
\frac{d\phi}{dr}\Bigr\rvert_{R_o}=\frac{\hbar I_o}{4 \pi \rho R_o}, \ \ \frac{d\phi}{dr}\Bigr\rvert_{R_i}=-\frac{\hbar I_i}{4 \pi \rho R_i}.
\ee

The Neumann boundary conditions at inner and outer edges are sufficient to select a unique solution to the 
second order differential equation for $\phi(r)$.  
For some values of the currents at the inner and outer edges, $I_i,I_o$, there are no solutions to Eq.~\eqref{RadialSG}.
Whenever this is true, the condensate must be time-dependent. 
Figure~\ref{Ii-Io-corbino} plots, for several value of $\lambda$,
the boundary of the region in the $I_o-I_i$ plane over which the condensate is able to reach a time independent steady state
for several different values of $\lambda$.  When currents are expressed in units of the macrospin critical current $I_c$, 
the stability region is largest for the largest values of $\lambda$. We see from Figure~\ref{Ii-Io-corbino} that 
the macrospin limit is accurately approached already at $\lambda \sim R$.
Since $\lambda$ is reduced by increasing the strength of interlayer tunneling $\Delta_{SAS}$, 
a change that can be experimentally implemented
by slightly reducing the interlayer barrier thickness, it should be 
possible to test the deviations from the criterion $|I_o+I_i|<I_c$ experimentally.
Deviations from the simple macrospin stability criterion, $|I_o+I_i|<I_c$, 
will signal the departure from the macrospin limit as illustrated in Fig.~\ref{Ii-Io-corbino}. 
Measurements of this type combined with detailed modeling of the sample geometry can be used to experimentally estimate the condensate stiffness $\rho$.

\section{Time-Dependent Transport Characteristics}
\label{sec:AC}

The tunneling I-V characteristics described so far were obtained by averaging the precession rate of the condensate phase and the transport 
currents over time.  They therefore correspond to {\it d.c.} measurements. 
However, above the critical bias where the condensate phase starts to precess the current can have an a.c. component in response to a
 d.c. bias.  Therefore a.c. current measurements can provide additional information on the condensate dynamics. 

\begin{figure}
\includegraphics[width=3.2in]{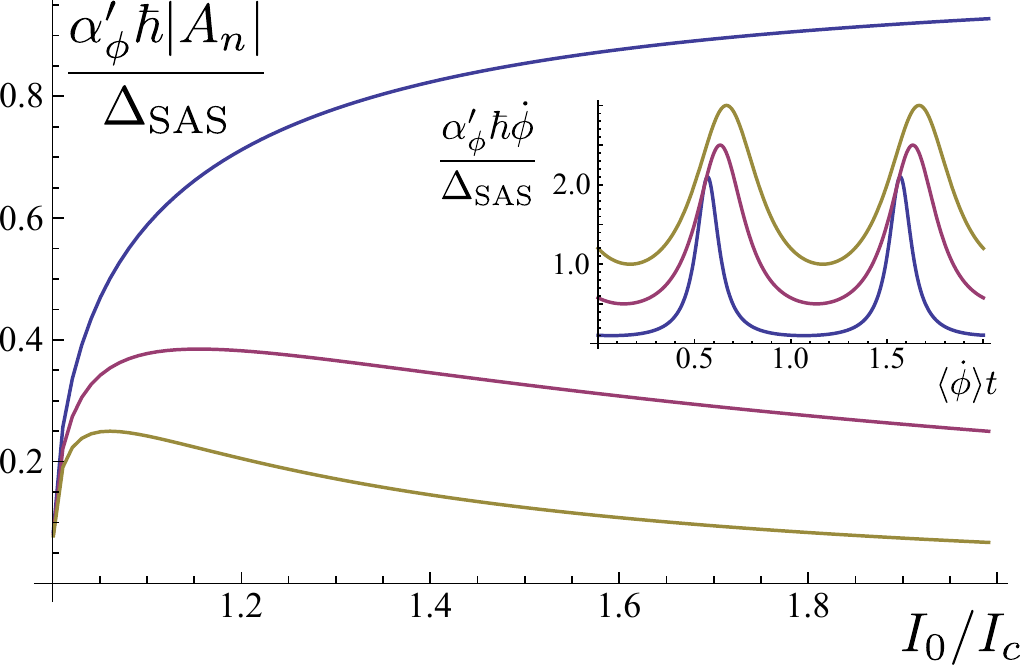}
\caption{In the time-dependent regime of a tunneling experiment the periodic precession rate of the condensate phase leads to a periodic
contribution to the interlayer current.  This figure plots harmonics of the precession rate of the condensate as a function $\mu/\mu_c$, which can be measured as an a.c. voltage in the inner rim of the Corbino annulus. Here $A^0_1\equiv g\mu_c/(g_{qp}+g)$.}
\label{harmonics}
\end{figure}

Equation~\eqref{eq:macrospin} can be analytically solved for the instantaneous condensate precession rate.
When the bias exceeds the critical value, $\mu>\mu_c\equiv\hbar I_c/g$:

\be
\hbar \dot{\phi}=\left(\frac{g \mu}{g_{qp}+g}\right)\frac{(\mu^2-\mu_c^2) \sec^2(\langle\dot{\phi} \rangle t/2)}{\mu^2+[\mu_c+\sqrt{\mu^2-\mu_c^2} \tan(\langle\dot{\phi} \rangle t/2)]^2}
\ee  


\noindent where $\langle\dot{\phi} \rangle$ is given by Eq.~\eqref{phidot_av}. For biases that barely exceed the critical value, $\mu \gtrsim \mu_c$, $\dot{\phi}$ oscillates in a very anharmonic fashion. Figure~\ref{harmonics} illustrates the amplitudes, $|A_n|$, of the leading harmonics of the condensate precession rate, {\it i.e.} $\hbar \dot{\phi}=A_n e^{in \langle\dot{\phi}\rangle t}$, with the fundamental frequency given by the average precession rate $\langle\dot\phi\rangle$ from Eq.~\eqref{phidot_av}. Interestingly, in the limit $\mu \gg \mu_c$ all higher harmonics die out, $A_n\rightarrow 0$ for $n\geq2$, but the fundamental harmonic remains finite with an amplitude given by: $A_1\rightarrow A^0_1\equiv g\mu_c/(g_{qp}+g)$. Thus the rate of precession becomes sinusoidal: 

\be
\hbar \dot{\phi} \approx \frac{g}{g+g_{qp}}\left[\mu-  \mu_c\cos (\langle \dot{\phi}\rangle t)\right], \ \ \mu \gg \mu_c.
\ee

\begin{figure}
\includegraphics[width=3in]{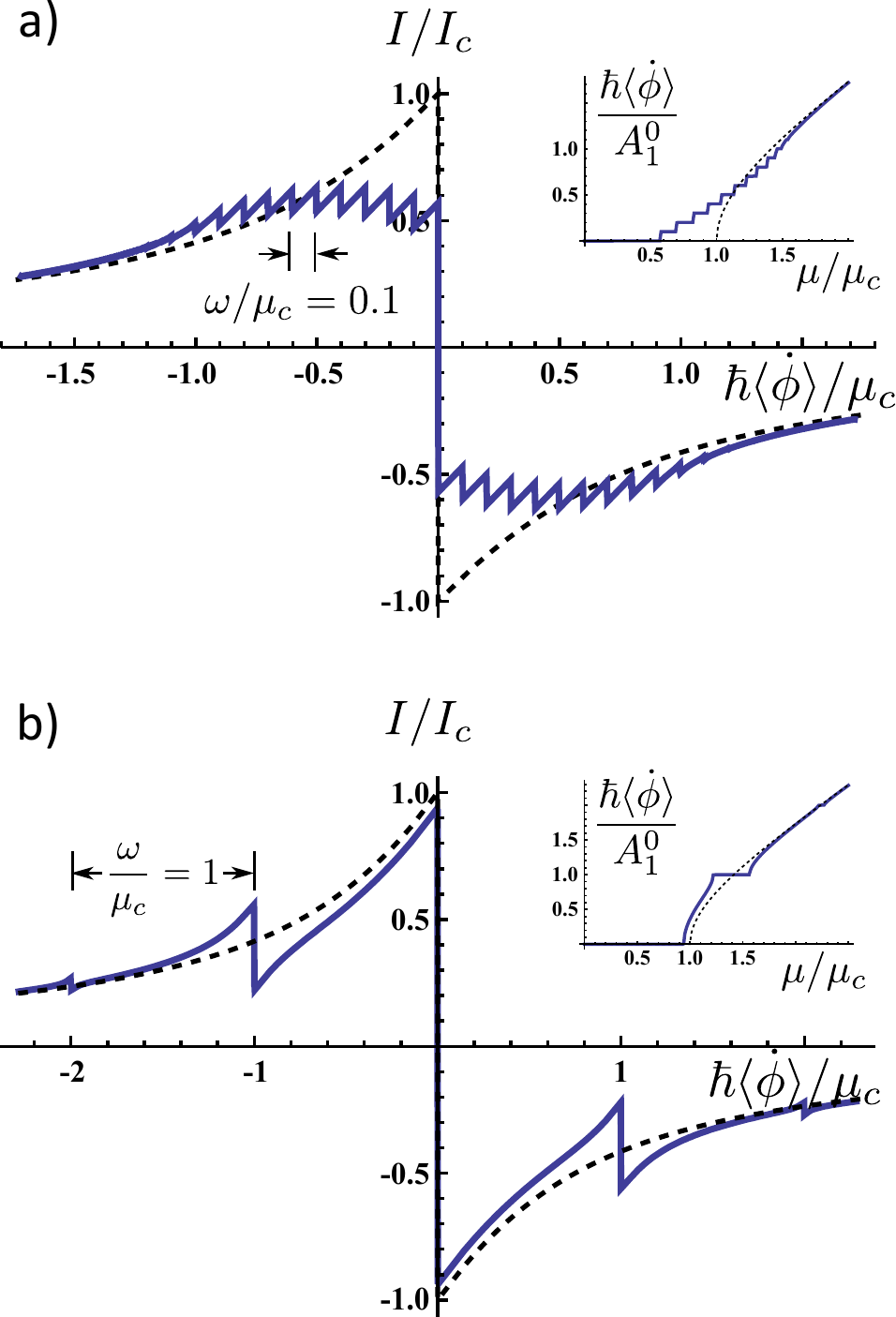}
\caption{(color online) Illustration of the Shapiro steps that arise when an a.c. signal is added to the d.c. bias. Panel a) corresponds to a driving frequency $\hbar \omega=0.1 \mu_c$, and panel b) to $\hbar \omega=\mu_c$. In both cases the amplitude of the oscillating bias was taken to be $\delta \mu=\mu_c/2$. The insets illustrate the behavior of the averaged precession rate of the condensate, which is proportial to the tunneling voltage measured in the inner edge of the Corbino ring. The dashed black lines correspond to the behavior for pure d.c. bias illustrated in Fig.~\ref{IVs}. We chose $g_{qp}\rightarrow 0$.}
\label{Shapiro}
\end{figure}

Another interesting case is that of driving voltages with a finite a.c. component: $\mu+\delta \mu \sin(\omega t)$.  With an applied bias
of this form the I-V characteristics display the analog of the Shapiro steps~\cite{Hyart2013} that are observed in Josephson junctions~\cite{Grimes1968}. 
In this regime, the average precession rate will lock at discrete values which are integer multiples of the driving frequency, {\it i.e.} $\langle\dot{\phi}\rangle=k \omega$ with $k\in \mathbb{Z}$ over finite intervals of the average driving voltage $\mu$ as illustrated in the insets of Figs.~\ref{Shapiro}(a)-(b). 
Consequently the average d.c. voltage measured at the inner rim of the Corbino annulus will remain fixed at discrete values $- k \hbar \omega$, over finite ranges of the average driving voltage $\mu$. The width of the $k$-th step, $\Delta \mu_k$, is controlled by the amplitude of a.c. component of the driving voltage $\delta \mu$. Approximate expressions for the values of the widths of the first steps in the limit of slow driving frequencies ($\hbar \omega\ll \mu_c$) and
small driving amplitudes ($\delta \mu \ll \mu_c$), derived in Ref.~\onlinecite{Renne1974}, are listed in Table~\ref{widths}.

\begin{table}
\caption{Widths of the Shapiro steps in the limit of slow driving frequencies and small driving amplitudes.} 
\centering 
\begin{tabular}{|c | c |} 
 \hline
$k\in \mathbb{Z}$  & $\Delta \mu_k /\mu_c$ \\ 
 \hline
$0$  & $2-2 \frac{\delta \mu}{\mu_c}+2 \left(1+\frac{g_{qp}}{g}\right)\frac{\hbar\omega}{\mu_c}\sqrt{\frac{\delta \mu}{\mu_c}}$ \\ 

$1$  & $2 \left(1+\frac{g_{qp}}{g}\right) \frac{\hbar \omega}{\mu_c}\sqrt{\frac{\delta \mu}{\mu_c}}$ \\ 

$2$  & $2 \left(1+\frac{g_{qp}}{g}\right) \frac{\hbar \omega}{\mu_c}\sqrt{\frac{\delta \mu}{\mu_c}}$ \\ [1ex] 
\hline
\end{tabular}
\label{widths} 
\end{table}

The widths of the steps in the opposite limit of fast driving frequencies, $\hbar \omega\gg \mu_c, \delta \mu$, can be shown to be given by:

\be
\Delta \mu_k = 2 \ \mu_c \ J_k \left[\frac{\delta \mu}{\hbar \omega (1+g_{qp}/g)}\right],
\ee

\noindent with $J_k$ the $k$-th Bessel function~\cite{Grimes1968,Renne1974}. 

The critical currents in recent samples are on the order of $e I_c \sim 1$nA~\cite{Huang2012,Nandi2013}. To realize discernible steps it is desirable that the applied bias frequency reaches values no more than 
an order of magnitude smaller than $I_c$. This means that the
desired frequencies are $\omega \gtrsim 10$GHz should be attainable by pure electronic means. 
It might also be interesting to explore the coupling of the condensate to microwave radiation.  
Note that the predictions made here for Shapiro step widths differ qualitatively from those 
in Ref.~\onlinecite{Hyart2013}, allowing future experiments to distinguish between
the two different theoretical frameworks.  

\section{Impact of disorder}\label{Sec:disc}

We have demonstrated that the assumption of a long-ranged nearly uniform order in coherent quantum Hall bilayers is able to explain
all the surprising transport anomalies that this system displays.  We have so far adopted a mainly 
phenomenological viewpoint and have not addressed in detail the microscopic physics which 
determines the values of the coupling constants in the order parameter energy functional. We now propose a physical 
interpretation of the phase variable in our theory
which  extends previous arguments~\cite{Su2010,Pesin2011} that disorder 
drastically reduces the strength of the coupling of interlayer tunneling to the condensate.    

\begin{figure}
\includegraphics[width=3.5in]{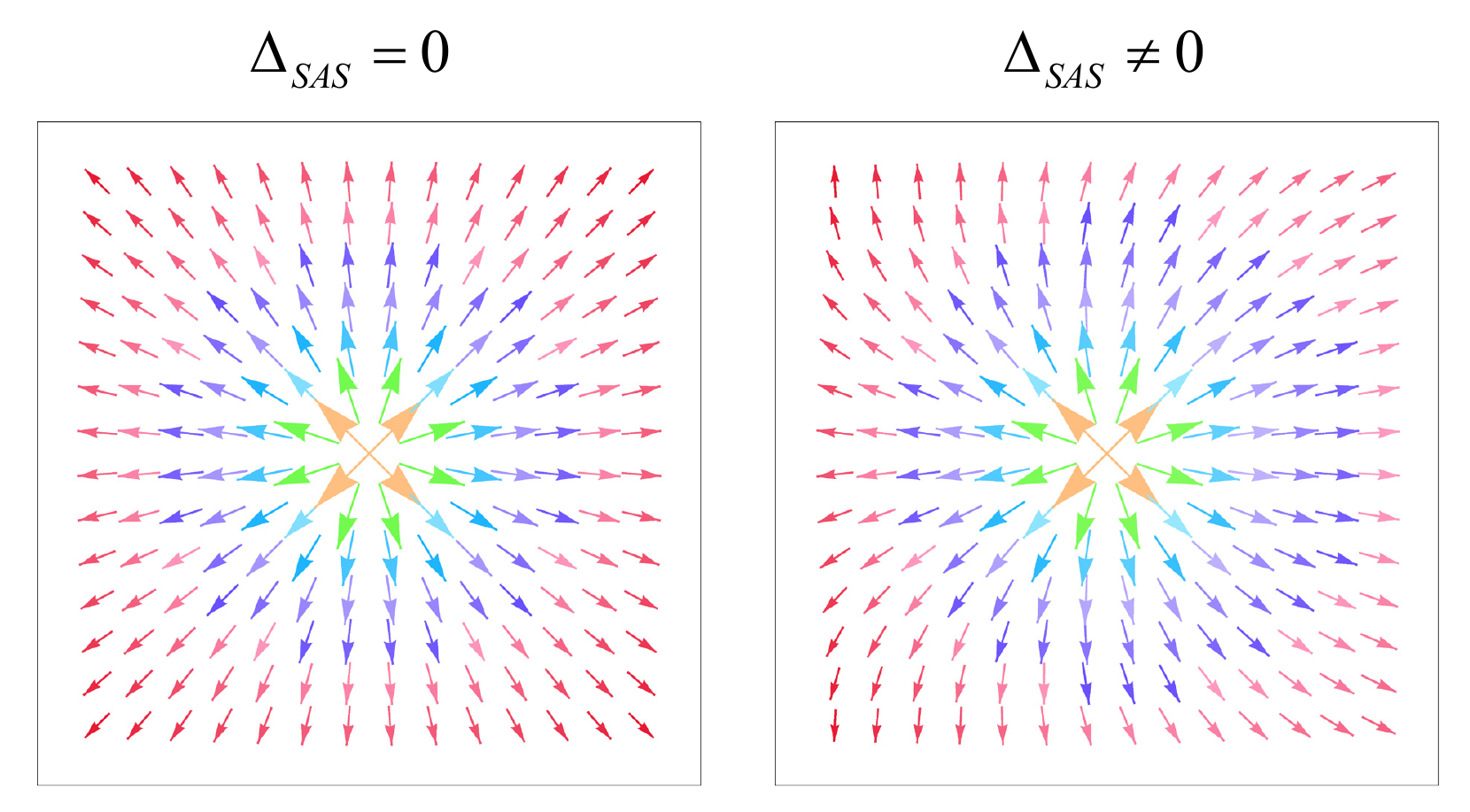}
\caption{(color online) Illustration of the impact of interlayer tunneling near a meron core. The right panel depicts how the presence of a weak tunneling term, which breaks the XY symmetry, produces a well defined spatially averaged phase.}
\label{disorder}
\end{figure}

Consider first the case of a bilayer with purely spontaneous interlayer phase coherence 
in the absence of interlayer tunneling.  
The strength of order in such a system is reduced by quantum and thermal fluctuations.
These effects have been addressed by many researchers, but are ignored below because their impact is~\cite{Su2010,Pesin2011} insufficient to 
explain the reduction in the strength of interlayer tunnel coupling to the condensate by four to five orders of magnitude found in experiments~\cite{Eisenstein2014}.  We concentrate instead on the influence of disorder, which couples to the interlayer phase 
because of the relationship~\cite{Sondhi1993,Moon1995} in quantum Hall systems between pseudospin textures
and charge density.   Smooth local density variations $n(\vec{r})$ are generated in quantum Hall ferromagnets when the spatial
gradients of both the interlayer phase ($\varphi$) and the layer polarization ($M_z$) are non-zero~\cite{Sondhi1993,Moon1995}:
\begin{equation} 
\delta n(\vec{r}) = -\frac{1}{4\pi} \hat{M}(\vec{r}) \cdot [ \partial_x \hat{M}(\vec{r}) \times \partial_y \hat{M}(\vec{r})]. 
\end{equation}
Smooth random potential 
variation, which is always present in high-mobility two-dimensional electron systems, therefore favors pseudospin textures and 
can~\cite{Moon1995} drive the formation of vortices and anti-vortices (merons) in the pseudospin order parameter. 
Current experiments can be understood under the assumption that disorder in the absence of interlayer tunneling is 
sufficient to completely destroy long range order in the interlayer phase.  This is a natural assumption because of the relationship between disorder potentials and pseudospin textures.

We refer to the interlayer phase configuration 
which minimizes the sum of the disorder energy,
\begin{equation}
E_{dis} = \int d^2\vec{r} \, n(\vec{r}) \, V_{dis}(\vec{r}), 
\end{equation} 
and the pseudospin stiffness energy,
\begin{equation} 
E_{rho} =  \rho  \int d^2 \vec{r} \,\;  |\nabla \varphi(\vec{r})|^2,
\end{equation} 
as $\varphi_0(\vec{r})$.  In writing the gradient energy in this form 
we are assuming that the pseudospin lies in its $\hat{x}-\hat{y}$ plane 
with local azimuthal orientation $\varphi(\vec{r})$ except near small 
meron (vortex) cores. Because long-range phase order is absent in the absence of inter-layer tunneling 
we have: 
\begin{eqnarray} 
\langle \cos(\varphi_{0}(\vec{r})) \rangle_{\vec{r}} &=& 0,  \nonumber \\
\langle \sin(\varphi_{0}(\vec{r})) \rangle_{\vec{r}} &=& 0,   \nonumber \\
\langle \cos^2(\varphi_{0}(\vec{r})) \rangle_{\vec{r}} &=& 1/2,  \nonumber \\
\langle \sin^2(\varphi_{0}(\vec{r})) \rangle_{\vec{r}} &=& 1/2.
\end{eqnarray}
where the angle brackets denote spatial averages.  
Because only phase gradients yield charge densities, disorder determines the interlayer phase only up to a global constant.
The characteristic length scale for variations in $\varphi_0(\vec{r})$ is the disorder correlation length $\xi$.  

Now consider how the ground state phase configuration is altered when interlayer tunneling is present.  Because the interlayer tunneling explcitly breaks the global XY symmetry, the system can develop a global phase even after spatially averaging, as illustrated in Fig.~\ref{disorder}. The interlayer tunneling 
energy is: 
\begin{equation} 
\label{eq:etun} 
E_{tun} = -n \Delta_{SAS}  \int d^2\vec{r} \cos(\varphi(\vec{r})). 
\end{equation} 
In the pseudospin language, the interlayer tunneling produces a spatially constant effective magnetic field
which acts in the $\hat{x}$ direction and has magnitude $\Delta_{SAS}$.  In the same language
$\varphi_0(\vec{r})$ describes the orientation of a pseudospin that is aligned with a spatially
varying pseudospin effective magnetic field with a typical magnitude $ \sim \rho / n \xi^2$, which is much larger than $\Delta_{SAS}$.  Here 
$n = (2 \pi \ell^2)^{-1}$ is the full Landau level density and the factor of $n$ in the denominator renormalizes
energy per area to energy per electron.  The additional weak field will result in a shifted ground 
state interlayer phase, 
\begin{equation}
\varphi_{GS}(\vec{r}) = \varphi_0(\vec{r}) + \delta \varphi(\vec{r}),
\end{equation} 
where $\delta \varphi(\vec{r})$ is describes the reorientation of
pseudo-spins towards the $\hat{x}$ direction.  The phase shift is largest in magnitude when the unperturbed pseudospin is 
perpendicular to the $\hat{x}$ direction, and 
has a typical magnitude proportional to the ratio of the 
two fields.  It follows that $\delta \varphi \ll 1$, and that: 
\begin{equation} 
\label{eq:deltaphi}
\delta \varphi(\vec{r}) \sim - \frac{\Delta_{SAS} n \xi^2 \, \sin(\varphi_0(\vec{r}))}{\rho}, 
\end{equation} 
and hence that:
\begin{eqnarray} 
\langle \cos(\varphi_{GS}(\vec{r})) \rangle_{\vec{r}}  \; &\sim& \frac{\Delta_{SAS} n \xi^2 }{2 \rho}, \nonumber \\
\langle \sin(\varphi_{GS}(\vec{r})) \rangle_{\vec{r}}    \; &=& \;   0.  
\end{eqnarray}

Let us now consider the influence of the transport bias voltages and 
the pseudospin transport torques. In the presence of transport currents the condensate phase is given by: 
\begin{equation} 
\varphi(\vec{r}) \to \varphi_{GS}(\vec{r}) + \phi(\vec{r}),
\end{equation} 
\noindent where $\phi(\vec{r})$ describes the local deviation from the equilibrium orientation in the presence of the tunneling term. We will argue that $\phi(\vec{r})$ should be identified with the phase variable employed in the main body of this paper and that it is very slowly varying so that it does not contribute significantly to the pseudospin-texture related charge density. 
In particular, the dependence of total energy on $\phi(\vec{r})$ can be found to be given:

\begin{equation} 
\label{coarsegrain} 
E[\phi] \approx E_{GS} - \int d^2\vec{r} \ [ \Delta  \cos(\phi(\vec{r}))  
+  \rho |\nabla \phi(\vec{r})|^2], 
\end{equation}

\noindent with:

\begin{equation} 
\label{coarsedelta} 
\Delta = \Delta_{SAS} n  \langle \cos(\varphi_{GS}(\vec{r})) \rangle_{\vec{r}} \sim \frac{(n \xi \Delta_{SAS})^2 }{2 \rho}.  
\end{equation}

\noindent Equation~\eqref{coarsegrain} applies provided that $\phi$ varies slowly on the disorder length scale $\xi$ and 
is based on the observation that  $\langle \nabla \varphi_{GS}(\vec{r}) \rangle_{\vec{r}} = 0$ because 
there is no macroscopic exciton flow in the absence of electrical bias voltages.   The final conclusion is that 
in Eq.~\eqref{eq:3} the effective stiffness parameter $\rho$ is not strongly influenced by disorder, but the the effective interlayer tunneling parameter is strongly suppressed from $n \Delta_{SAS}$ to $\Delta \sim (n \xi \Delta_{SAS})^2/(2 \rho)$. 

Let us now illustrate how this picture is in quantitative agreement with experiments. In the early experiments of Spielman {\it et al.}, the typical value of the 
tunneling amplitude was estimated to be $\Delta_{SAS} \sim 10^{-9} {\rm eV}$~\cite{Spielman2004}. 
The stiffness energy scale at $d=0$ is 
$\rho=(e^2/\epsilon l)/(32\sqrt{2\pi})\sim 10^{-4}$eV~\cite{Sondhi1993,Moon1995}. 
In mean-field theory, the stiffness is reduced by a further factor of $\sim 3$ at the values of 
$d/\ell$ studied experimentally and quantum fluctuations~\cite{Joglekar2001} can reduce the stiffness
further. 
The macrospin critical current one would naively expect in the absence of disorder for these samples, which
have an area $A\sim 250 \times 250 \mu$m$^2$, is about $e I_c \sim 10^{-7} A$. 
However, these experiments found a critical current about four orders of 
magnitude smaller: $e I_c \sim 10^{-11}$A~\cite{Spielman2004}. 
The relative smallness of this critical current is a signature of the strongly reduced effective tunneling amplitude. Estimating the impact of disorder requires knowledge of the typical values for the disorder correlation length. 
As we argue below, in-plane field measurements of the critical current are consistent with a value of the disorder correlation length of about $n\xi^2 \sim 10$. 
With this estimate one obtains that the critical current is expected to be reduced to about $e I_c \sim 10^{-11}$A, in agreement with experiment. Furthermore, using paramters from more recent experiments with larger tunneling amplitudes of about $\Delta_{SAS} \sim 10^{-8} {\rm eV}$~\cite{Tiemann2009}, the mechanism described in this section predicts an enhanced critical current of about $e I_c\sim 10^{-9}$A, which is again in agreement with their findings. We therefore conclude that this mechanism is the main factor behind the dramatic reduction of the critical current.  

The property that the critical current is proportional to the square of the microscopic interlayer tunneling amplitude $\Delta_{SAS}$ is consistent with experiment, including very recent measurements~\cite{Zhang2014} which
compared similar samples with different tunnel-barrier thicknesses,  although very 
precise comparisons require measurements at current values close to the 
critical value and are therefore challenged by circuit 
instabilities.  Uncontrolled wafer-scale spatial variation in the tunnel barrier thickness
can~\cite{Eisenstein2014} also complicate comparison between theory and experiment.

We would like now to discuss the implications of this picture for in-plane field measurements. In pseudospin language, an in-plane magnetic field causes the 
effective magnetic field due to tunneling to precess around the $\hat z$ axis, instead of being fixed in
the $\hat{x}$ direction.  The azimuthal orientation of the pseudo spin field is~\cite{Moon1995} $\phi_{B} = Q x$ where 
$Q = 2 \pi B_\Vert d/\Phi_0$, $d$ is the layer separation, $\Phi_0$ is the magnetic flux quantum, and $x$ is the 
planar coordinate perpendicular to the in-plane magnetic field direction.  
If the macrospin limit were achieved in the absence of disorder,  the condensate 
stiffness would suppress its phase response to the precessing in-plane field.
The condensate would then couple only to the spatial average
average of the in-plane field, yielding a Fraunhofer pattern analogous to that found 
in Josephson junctions.  
For $QL$ large compared to system-size, the critical current should be 
strongly reduced.  In contrast the experimental finding is that the critical current is reduced only 
gradually and at a field scale corresponding to 
$Q^{-1} \sim 0.1 {\rm \mu m}$~\cite{Spielman2001,Spielman2004,Zhang2014}, which is 
smaller than the sample sizes typically used in bilayer exciton condensate studies.  
The picture outlined above also provides a simple explanation for the unexpectedly weak dependence of 
tunneling transport anomalies on 
in-plane magnetic fields.  Instead of being reoriented toward the $\hat{x}$ direction by tunneling, the pseudospin will be 
reoriented toward the local effective field direction.  Eq.~\eqref{eq:etun} is therefore modified to: 
\begin{equation} 
\label{eq:etunB} 
E_{tun} = - \frac{\Delta_{SAS}}{2\pi\ell^2}  \int d^2\vec{r} \cos(\varphi(\vec{r})-Qx), 
\end{equation} 
and Eq.~\eqref{eq:deltaphi} to: 
\begin{equation} 
\delta \varphi(\vec{r}) \sim - \frac{\Delta_{SAS} n \xi^2 \, \sin(\varphi_0(\vec{r})-Qx)}{\rho}, 
\end{equation} 
but Eq.~\eqref{coarsegrain} is unaltered.  Changes are expected in an in-plane magnetic field 
only when $Q \xi \gtrsim 1$.  In-plane field dependent 
measurements therefore allow one to extract a value for the disorder correlation length of about $\xi\sim 0.1\mu$m. 

The ideas discussed in this paper can be tested experimentally by increasing the bare microscopic 
interlayer tunneling amplitude sufficiently to escape from the macrospin limit.  The length-scale $\lambda=\sqrt{2 \rho/\Delta}$ over which variations of the coarse grained phase $\phi$ occur, can be estimated by using the experimentally measured critical currents to be about $\sim 1mm$ for the samples described in Ref.~\cite{Spielman2004}, and about $\sim 0.1mm$ in the samples of Ref.~\cite{Tiemann2009}. In the presence of disorder one expects $\lambda$ to be inversely proportional to the bare tunneling amplitude, $\lambda \sim2 \rho/(n \xi \Delta_{SAS})$. Because $\Delta_{SAS}$ is exponentially sensitive to both the height ({\it i.e.} to the Al fraction) and the width of the tunnel barrier between quantum wells, there does not seem to be a serious obstacle to increasing the critical current by increasing $\Delta_{SAS}$, while reducing $\lambda$. By increasing $\Delta_{SAS}$ by about an order of magnitude with respect to current values reported in Refs.~\cite{Tiemann2009,Nandi2013} one should be able to attain the conditions in which the deviations from the macrospin limit illustrated in Fig.~\ref{Ii-Io-corbino}. are observable.  

\section{Summary and discussion}

We have demonstrated that a satisfactory understanding of all transport anomalies associated with the 
exciton condensate state of quantum Hall bilayers~\cite{Eisenstein2014} can be achieved by combining a fermion mean-field 
quasiparticle theory of transport with a theory of the condensate which accounts for its response~\cite{Su2008} 
to quasiparticle excitonic Andreev scattering. This theory is closely related to the picture successfully used to describe magneto-electronic effects in circuits containing magnetic metals. It uses the same framework to explain anomalous interlayer tunneling, 
quantized Hall drag, the absence of a Hall voltage in Hall bar counterflow measurements, 
and perfect Coulomb drag and non-local coupling between tunneling at inner and outer
edges in Corbino geometry bilayers.  Using this theory we have shown that recent 
measurements in Corbino devices~\cite{Huang2012,Nandi2013} 
directly demonstrate that samples studied to date are in a macrospin limit
in which the condensate stiffness is sufficient to maintain a spatially constant value of the 
coarse grained condensate phase.  The constant phase is 
unexpected because interlayer tunneling strongly favors interlayer phases which return to zero 
away from current source and drain contacts over a length scale that is naively expected 
to be much smaller than the system size.
We have argued, however, that the phase that enters in the macrospin model must be interpreted as a coarse grained 
average of a local interlayer phase that is strongly scrambled by
disorder because of the relationship~\cite{Sondhi1993,Moon1995} between charge densities and pseudospin
textures in the quantum Hall regime.  
Using this idea we can explain a strong reduction of the effective interlayer tunneling strength by 
about four orders of magnitude from the naive value extracted from the bare tunneling amplitude: the reduction explains the applicability of the macrospin limit, the small 
experimental values of the tunneling anomaly critical currents, and the 
surprising weak dependence of tunneling on in-plane magnetic field.
We have applied the theory to predict  
anomalies in {\em ac} transport measurements, and to predict changes 
in the non-local relationship between tunneling characteristics at inner and 
outer Corbino sample edges in devices with stronger interlayer tunneling 
than has been studied to date.  Specifically we predict that 
the experimental finding that the critical current for the tunneling anomaly
is the sum of the currents at the two Corbino edges will break down in
samples with stronger interlayer tunneling~\cite{Eisenstein2014}.  

Our main goal in this paper is to lay out a theoretical framework for treating the 
transport properties of bilayer exciton condensate. In order to focus on this goal we have neglected some details which do play an experimental role.  In particular we have 
concentrated on the zero-temperature limit by assuming a perfect quantum 
Hall effect and by neglecting thermal fluctuations of the condensate.  
Only when the quantum Hall effect is perfect is our assumption that quasiparticles  
travel ballistically between source, drain, and voltage contacts on the same 
sample edge - and never between contacts on opposite edges - exactly valid.  
The property that a perfect quantum Hall effect is approached as $T \to 0$ 
makes quasiparticle transport in this limit very simple and allows our theoretical 
framework to be directly tested. However, the theory is very readily generalized to account for an imperfect quantum Hall effect. Similarly thermal condensate fluctuations can readily be accounted  for our theory by adding a stochastic contribution to the condensate equation of motion that is consistent with the fluctuation-dissipation theorem and with the model used for dissipative incoherent quasiparticle tunneling between layers.  Finally, there is good reason to expect quasiparticle-tunneling to be both 
non-Ohmic and temperature-dependent.   The Ohmic tunneling assumption is adopted 
here for illustrative purposes only.     

Achieving a complete understanding of transport in quantum Hall bilayers is not only of fundamental interest, 
but also useful for gaining insight into transport in systems that at first glance appear to be
quite different but which share crucial similarities. 
Indeed, we have recently exploited the similarity between coherent bilayers and easy-plane 
ferromagnets to propose a new kind of spintronic device that displays non-local coupling between spatially separated
transport channels~\cite{Chen2014} that is analogous to that displayed by the Corbino geometry samples explored
in this paper. Understanding similarities and differences between these kind of systems might prove
useful in advancing both fundamental and applied condensed matter physics research.

\begin{acknowledgements}
We are grateful to W. Dietsche, J.P. Eisenstein, B.I. Halperin, K. von Klitzing, and D. Zhang 
for valuable interactions.   This work was supported by DOE Division of Materials Sciences and Engineering
grant DE-FG03-02ER45958 and Welch Foundation grant TBF1473.
I.S. acknowledges the support of the Pappalardo Fellowship in Physics.
\end{acknowledgements}


\begin{thebibliography}{66}%
\makeatletter
\providecommand \@ifxundefined [1]{%
 \@ifx{#1\undefined}
}%
\providecommand \@ifnum [1]{%
 \ifnum #1\expandafter \@firstoftwo
 \else \expandafter \@secondoftwo
 \fi
}%
\providecommand \@ifx [1]{%
 \ifx #1\expandafter \@firstoftwo
 \else \expandafter \@secondoftwo
 \fi
}%
\providecommand \natexlab [1]{#1}%
\providecommand \enquote  [1]{``#1''}%
\providecommand \bibnamefont  [1]{#1}%
\providecommand \bibfnamefont [1]{#1}%
\providecommand \citenamefont [1]{#1}%
\providecommand \href@noop [0]{\@secondoftwo}%
\providecommand \href [0]{\begingroup \@sanitize@url \@href}%
\providecommand \@href[1]{\@@startlink{#1}\@@href}%
\providecommand \@@href[1]{\endgroup#1\@@endlink}%
\providecommand \@sanitize@url [0]{\catcode `\\12\catcode `\$12\catcode
  `\&12\catcode `\#12\catcode `\^12\catcode `\_12\catcode `\%12\relax}%
\providecommand \@@startlink[1]{}%
\providecommand \@@endlink[0]{}%
\providecommand \url  [0]{\begingroup\@sanitize@url \@url }%
\providecommand \@url [1]{\endgroup\@href {#1}{\urlprefix }}%
\providecommand \urlprefix  [0]{URL }%
\providecommand \Eprint [0]{\href }%
\providecommand \doibase [0]{http://dx.doi.org/}%
\providecommand \selectlanguage [0]{\@gobble}%
\providecommand \bibinfo  [0]{\@secondoftwo}%
\providecommand \bibfield  [0]{\@secondoftwo}%
\providecommand \translation [1]{[#1]}%
\providecommand \BibitemOpen [0]{}%
\providecommand \bibitemStop [0]{}%
\providecommand \bibitemNoStop [0]{.\EOS\space}%
\providecommand \EOS [0]{\spacefactor3000\relax}%
\providecommand \BibitemShut  [1]{\csname bibitem#1\endcsname}%
\let\auto@bib@innerbib\@empty
\bibitem [{\citenamefont {Spielman}\ \emph {et~al.}(2000)\citenamefont
  {Spielman}, \citenamefont {Eisenstein}, \citenamefont {Pfeiffer},\ and\
  \citenamefont {West}}]{Spielman2000}%
  \BibitemOpen
  \bibfield  {author} {\bibinfo {author} {\bibfnamefont {I.~B.}\ \bibnamefont
  {Spielman}}, \bibinfo {author} {\bibfnamefont {J.~P.}\ \bibnamefont
  {Eisenstein}}, \bibinfo {author} {\bibfnamefont {L.~N.}\ \bibnamefont
  {Pfeiffer}}, \ and\ \bibinfo {author} {\bibfnamefont {K.~W.}\ \bibnamefont
  {West}},\ }\href {\doibase 10.1103/PhysRevLett.84.5808} {\bibfield  {journal}
  {\bibinfo  {journal} {Phys. Rev. Lett.}\ }\textbf {\bibinfo {volume} {84}},\
  \bibinfo {pages} {5808} (\bibinfo {year} {2000})}\BibitemShut {NoStop}%
\bibitem [{\citenamefont {Eisenstein}\ and\ \citenamefont
  {MacDonald}(2004)}]{Eisenstein2004}%
  \BibitemOpen
  \bibfield  {author} {\bibinfo {author} {\bibfnamefont {J.}~\bibnamefont
  {Eisenstein}}\ and\ \bibinfo {author} {\bibfnamefont {A.}~\bibnamefont
  {MacDonald}},\ }\href@noop {} {\bibfield  {journal} {\bibinfo  {journal}
  {Nature}\ }\textbf {\bibinfo {volume} {432}},\ \bibinfo {pages} {691}
  (\bibinfo {year} {2004})}\BibitemShut {NoStop}%
\bibitem [{\citenamefont {Eisenstein}(2014)}]{Eisenstein2014}%
  \BibitemOpen
  \bibfield  {author} {\bibinfo {author} {\bibfnamefont {J.}~\bibnamefont
  {Eisenstein}},\ }\href {\doibase 10.1146/annurev-conmatphys-031113-133832}
  {\bibfield  {journal} {\bibinfo  {journal} {Annual Review of Condensed Matter
  Physics}\ }\textbf {\bibinfo {volume} {5}},\ \bibinfo {pages} {159} (\bibinfo
  {year} {2014})}\BibitemShut {NoStop}%
\bibitem [{\citenamefont {Kellogg}\ \emph {et~al.}(2002)\citenamefont
  {Kellogg}, \citenamefont {Spielman}, \citenamefont {Eisenstein},
  \citenamefont {Pfeiffer},\ and\ \citenamefont {West}}]{Kellogg2002}%
  \BibitemOpen
  \bibfield  {author} {\bibinfo {author} {\bibfnamefont {M.}~\bibnamefont
  {Kellogg}}, \bibinfo {author} {\bibfnamefont {I.~B.}\ \bibnamefont
  {Spielman}}, \bibinfo {author} {\bibfnamefont {J.~P.}\ \bibnamefont
  {Eisenstein}}, \bibinfo {author} {\bibfnamefont {L.~N.}\ \bibnamefont
  {Pfeiffer}}, \ and\ \bibinfo {author} {\bibfnamefont {K.~W.}\ \bibnamefont
  {West}},\ }\href {\doibase 10.1103/PhysRevLett.88.126804} {\bibfield
  {journal} {\bibinfo  {journal} {Phys. Rev. Lett.}\ }\textbf {\bibinfo
  {volume} {88}},\ \bibinfo {pages} {126804} (\bibinfo {year}
  {2002})}\BibitemShut {NoStop}%
\bibitem [{\citenamefont {Kellogg}\ \emph {et~al.}(2004)\citenamefont
  {Kellogg}, \citenamefont {Eisenstein}, \citenamefont {Pfeiffer},\ and\
  \citenamefont {West}}]{Kellogg2004}%
  \BibitemOpen
  \bibfield  {author} {\bibinfo {author} {\bibfnamefont {M.}~\bibnamefont
  {Kellogg}}, \bibinfo {author} {\bibfnamefont {J.~P.}\ \bibnamefont
  {Eisenstein}}, \bibinfo {author} {\bibfnamefont {L.~N.}\ \bibnamefont
  {Pfeiffer}}, \ and\ \bibinfo {author} {\bibfnamefont {K.~W.}\ \bibnamefont
  {West}},\ }\href {\doibase 10.1103/PhysRevLett.93.036801} {\bibfield
  {journal} {\bibinfo  {journal} {Phys. Rev. Lett.}\ }\textbf {\bibinfo
  {volume} {93}},\ \bibinfo {pages} {036801} (\bibinfo {year}
  {2004})}\BibitemShut {NoStop}%
\bibitem [{\citenamefont {Tutuc}\ \emph {et~al.}(2004)\citenamefont {Tutuc},
  \citenamefont {Shayegan},\ and\ \citenamefont {Huse}}]{Tutuc2004}%
  \BibitemOpen
  \bibfield  {author} {\bibinfo {author} {\bibfnamefont {E.}~\bibnamefont
  {Tutuc}}, \bibinfo {author} {\bibfnamefont {M.}~\bibnamefont {Shayegan}}, \
  and\ \bibinfo {author} {\bibfnamefont {D.~A.}\ \bibnamefont {Huse}},\ }\href
  {\doibase 10.1103/PhysRevLett.93.036802} {\bibfield  {journal} {\bibinfo
  {journal} {Phys. Rev. Lett.}\ }\textbf {\bibinfo {volume} {93}},\ \bibinfo
  {pages} {036802} (\bibinfo {year} {2004})}\BibitemShut {NoStop}%
\bibitem [{\citenamefont {Wiersma}\ \emph {et~al.}(2004)\citenamefont
  {Wiersma}, \citenamefont {Lok}, \citenamefont {Kraus}, \citenamefont
  {Dietsche}, \citenamefont {von Klitzing}, \citenamefont {Schuh},
  \citenamefont {Bichler}, \citenamefont {Tranitz},\ and\ \citenamefont
  {Wegscheider}}]{Wiersma2004}%
  \BibitemOpen
  \bibfield  {author} {\bibinfo {author} {\bibfnamefont {R.~D.}\ \bibnamefont
  {Wiersma}}, \bibinfo {author} {\bibfnamefont {J.~G.~S.}\ \bibnamefont {Lok}},
  \bibinfo {author} {\bibfnamefont {S.}~\bibnamefont {Kraus}}, \bibinfo
  {author} {\bibfnamefont {W.}~\bibnamefont {Dietsche}}, \bibinfo {author}
  {\bibfnamefont {K.}~\bibnamefont {von Klitzing}}, \bibinfo {author}
  {\bibfnamefont {D.}~\bibnamefont {Schuh}}, \bibinfo {author} {\bibfnamefont
  {M.}~\bibnamefont {Bichler}}, \bibinfo {author} {\bibfnamefont {H.-P.}\
  \bibnamefont {Tranitz}}, \ and\ \bibinfo {author} {\bibfnamefont
  {W.}~\bibnamefont {Wegscheider}},\ }\href {\doibase
  10.1103/PhysRevLett.93.266805} {\bibfield  {journal} {\bibinfo  {journal}
  {Phys. Rev. Lett.}\ }\textbf {\bibinfo {volume} {93}},\ \bibinfo {pages}
  {266805} (\bibinfo {year} {2004})}\BibitemShut {NoStop}%
\bibitem [{\citenamefont {Finck}\ \emph {et~al.}(2011)\citenamefont {Finck},
  \citenamefont {Eisenstein}, \citenamefont {Pfeiffer},\ and\ \citenamefont
  {West}}]{Finck2011}%
  \BibitemOpen
  \bibfield  {author} {\bibinfo {author} {\bibfnamefont {A.~D.~K.}\
  \bibnamefont {Finck}}, \bibinfo {author} {\bibfnamefont {J.~P.}\ \bibnamefont
  {Eisenstein}}, \bibinfo {author} {\bibfnamefont {L.~N.}\ \bibnamefont
  {Pfeiffer}}, \ and\ \bibinfo {author} {\bibfnamefont {K.~W.}\ \bibnamefont
  {West}},\ }\href {\doibase 10.1103/PhysRevLett.106.236807} {\bibfield
  {journal} {\bibinfo  {journal} {Phys. Rev. Lett.}\ }\textbf {\bibinfo
  {volume} {106}},\ \bibinfo {pages} {236807} (\bibinfo {year}
  {2011})}\BibitemShut {NoStop}%
\bibitem [{\citenamefont {Nandi}\ \emph {et~al.}(2012)\citenamefont {Nandi},
  \citenamefont {Finck}, \citenamefont {Eisenstein}, \citenamefont {Pfeiffer},\
  and\ \citenamefont {West}}]{Nandi2012}%
  \BibitemOpen
  \bibfield  {author} {\bibinfo {author} {\bibfnamefont {D.}~\bibnamefont
  {Nandi}}, \bibinfo {author} {\bibfnamefont {A.}~\bibnamefont {Finck}},
  \bibinfo {author} {\bibfnamefont {J.}~\bibnamefont {Eisenstein}}, \bibinfo
  {author} {\bibfnamefont {L.}~\bibnamefont {Pfeiffer}}, \ and\ \bibinfo
  {author} {\bibfnamefont {K.}~\bibnamefont {West}},\ }\href@noop {} {\bibfield
   {journal} {\bibinfo  {journal} {Nature}\ }\textbf {\bibinfo {volume}
  {488}},\ \bibinfo {pages} {481} (\bibinfo {year} {2012})}\BibitemShut
  {NoStop}%
\bibitem [{\citenamefont {Fertig}(1989)}]{Fertig1989}%
  \BibitemOpen
  \bibfield  {author} {\bibinfo {author} {\bibfnamefont {H.~A.}\ \bibnamefont
  {Fertig}},\ }\href {\doibase 10.1103/PhysRevB.40.1087} {\bibfield  {journal}
  {\bibinfo  {journal} {Phys. Rev. B}\ }\textbf {\bibinfo {volume} {40}},\
  \bibinfo {pages} {1087} (\bibinfo {year} {1989})}\BibitemShut {NoStop}%
\bibitem [{\citenamefont {Eisenstein}\ \emph {et~al.}(1992)\citenamefont
  {Eisenstein}, \citenamefont {Boebinger}, \citenamefont {Pfeiffer},
  \citenamefont {West},\ and\ \citenamefont {He}}]{Eisenstein1992}%
  \BibitemOpen
  \bibfield  {author} {\bibinfo {author} {\bibfnamefont {J.~P.}\ \bibnamefont
  {Eisenstein}}, \bibinfo {author} {\bibfnamefont {G.~S.}\ \bibnamefont
  {Boebinger}}, \bibinfo {author} {\bibfnamefont {L.~N.}\ \bibnamefont
  {Pfeiffer}}, \bibinfo {author} {\bibfnamefont {K.~W.}\ \bibnamefont {West}},
  \ and\ \bibinfo {author} {\bibfnamefont {S.}~\bibnamefont {He}},\ }\href
  {\doibase 10.1103/PhysRevLett.68.1383} {\bibfield  {journal} {\bibinfo
  {journal} {Phys. Rev. Lett.}\ }\textbf {\bibinfo {volume} {68}},\ \bibinfo
  {pages} {1383} (\bibinfo {year} {1992})}\BibitemShut {NoStop}%
\bibitem [{\citenamefont {Murphy}\ \emph {et~al.}(1994)\citenamefont {Murphy},
  \citenamefont {Eisenstein}, \citenamefont {Boebinger}, \citenamefont
  {Pfeiffer},\ and\ \citenamefont {West}}]{Murphy1994}%
  \BibitemOpen
  \bibfield  {author} {\bibinfo {author} {\bibfnamefont {S.~Q.}\ \bibnamefont
  {Murphy}}, \bibinfo {author} {\bibfnamefont {J.~P.}\ \bibnamefont
  {Eisenstein}}, \bibinfo {author} {\bibfnamefont {G.~S.}\ \bibnamefont
  {Boebinger}}, \bibinfo {author} {\bibfnamefont {L.~N.}\ \bibnamefont
  {Pfeiffer}}, \ and\ \bibinfo {author} {\bibfnamefont {K.~W.}\ \bibnamefont
  {West}},\ }\href {\doibase 10.1103/PhysRevLett.72.728} {\bibfield  {journal}
  {\bibinfo  {journal} {Phys. Rev. Lett.}\ }\textbf {\bibinfo {volume} {72}},\
  \bibinfo {pages} {728} (\bibinfo {year} {1994})}\BibitemShut {NoStop}%
\bibitem [{\citenamefont {MacDonald}\ \emph {et~al.}(1990)\citenamefont
  {MacDonald}, \citenamefont {Platzman},\ and\ \citenamefont
  {Boebinger}}]{MacDonald1990}%
  \BibitemOpen
  \bibfield  {author} {\bibinfo {author} {\bibfnamefont {A.~H.}\ \bibnamefont
  {MacDonald}}, \bibinfo {author} {\bibfnamefont {P.~M.}\ \bibnamefont
  {Platzman}}, \ and\ \bibinfo {author} {\bibfnamefont {G.~S.}\ \bibnamefont
  {Boebinger}},\ }\href {\doibase 10.1103/PhysRevLett.65.775} {\bibfield
  {journal} {\bibinfo  {journal} {Phys. Rev. Lett.}\ }\textbf {\bibinfo
  {volume} {65}},\ \bibinfo {pages} {775} (\bibinfo {year} {1990})}\BibitemShut
  {NoStop}%
\bibitem [{\citenamefont {Brey}(1990)}]{Brey1990}%
  \BibitemOpen
  \bibfield  {author} {\bibinfo {author} {\bibfnamefont {L.}~\bibnamefont
  {Brey}},\ }\href {\doibase 10.1103/PhysRevLett.65.903} {\bibfield  {journal}
  {\bibinfo  {journal} {Phys. Rev. Lett.}\ }\textbf {\bibinfo {volume} {65}},\
  \bibinfo {pages} {903} (\bibinfo {year} {1990})}\BibitemShut {NoStop}%
\bibitem [{\citenamefont {Wen}\ and\ \citenamefont {Zee}(1992)}]{Wen1992}%
  \BibitemOpen
  \bibfield  {author} {\bibinfo {author} {\bibfnamefont {X.-G.}\ \bibnamefont
  {Wen}}\ and\ \bibinfo {author} {\bibfnamefont {A.}~\bibnamefont {Zee}},\
  }\href {\doibase 10.1103/PhysRevLett.69.1811} {\bibfield  {journal} {\bibinfo
   {journal} {Phys. Rev. Lett.}\ }\textbf {\bibinfo {volume} {69}},\ \bibinfo
  {pages} {1811} (\bibinfo {year} {1992})}\BibitemShut {NoStop}%
\bibitem [{\citenamefont {Wen}\ and\ \citenamefont {Zee}(1993)}]{Wen1993}%
  \BibitemOpen
  \bibfield  {author} {\bibinfo {author} {\bibfnamefont {X.~G.}\ \bibnamefont
  {Wen}}\ and\ \bibinfo {author} {\bibfnamefont {A.}~\bibnamefont {Zee}},\
  }\href {\doibase 10.1103/PhysRevB.47.2265} {\bibfield  {journal} {\bibinfo
  {journal} {Phys. Rev. B}\ }\textbf {\bibinfo {volume} {47}},\ \bibinfo
  {pages} {2265} (\bibinfo {year} {1993})}\BibitemShut {NoStop}%
\bibitem [{\citenamefont {Narikiyo}\ and\ \citenamefont
  {Yoshioka}(1993)}]{Narikiyo1993}%
  \BibitemOpen
  \bibfield  {author} {\bibinfo {author} {\bibfnamefont {O.}~\bibnamefont
  {Narikiyo}}\ and\ \bibinfo {author} {\bibfnamefont {D.}~\bibnamefont
  {Yoshioka}},\ }\href {\doibase 10.1143/JPSJ.62.1612} {\bibfield  {journal}
  {\bibinfo  {journal} {Journal of the Physical Society of Japan}\ }\textbf
  {\bibinfo {volume} {62}},\ \bibinfo {pages} {1612} (\bibinfo {year}
  {1993})}\BibitemShut {NoStop}%
\bibitem [{\citenamefont {Ezawa}\ and\ \citenamefont
  {Iwazaki}(1993)}]{Ezawa1993}%
  \BibitemOpen
  \bibfield  {author} {\bibinfo {author} {\bibfnamefont {Z.~F.}\ \bibnamefont
  {Ezawa}}\ and\ \bibinfo {author} {\bibfnamefont {A.}~\bibnamefont
  {Iwazaki}},\ }\href {\doibase 10.1103/PhysRevB.47.7295} {\bibfield  {journal}
  {\bibinfo  {journal} {Phys. Rev. B}\ }\textbf {\bibinfo {volume} {47}},\
  \bibinfo {pages} {7295} (\bibinfo {year} {1993})}\BibitemShut {NoStop}%
\bibitem [{\citenamefont {Yang}\ \emph {et~al.}(1994)\citenamefont {Yang},
  \citenamefont {Moon}, \citenamefont {Zheng}, \citenamefont {MacDonald},
  \citenamefont {Girvin}, \citenamefont {Yoshioka},\ and\ \citenamefont
  {Zhang}}]{Yang1994}%
  \BibitemOpen
  \bibfield  {author} {\bibinfo {author} {\bibfnamefont {K.}~\bibnamefont
  {Yang}}, \bibinfo {author} {\bibfnamefont {K.}~\bibnamefont {Moon}}, \bibinfo
  {author} {\bibfnamefont {L.}~\bibnamefont {Zheng}}, \bibinfo {author}
  {\bibfnamefont {A.~H.}\ \bibnamefont {MacDonald}}, \bibinfo {author}
  {\bibfnamefont {S.~M.}\ \bibnamefont {Girvin}}, \bibinfo {author}
  {\bibfnamefont {D.}~\bibnamefont {Yoshioka}}, \ and\ \bibinfo {author}
  {\bibfnamefont {S.-C.}\ \bibnamefont {Zhang}},\ }\href {\doibase
  10.1103/PhysRevLett.72.732} {\bibfield  {journal} {\bibinfo  {journal} {Phys.
  Rev. Lett.}\ }\textbf {\bibinfo {volume} {72}},\ \bibinfo {pages} {732}
  (\bibinfo {year} {1994})}\BibitemShut {NoStop}%
\bibitem [{\citenamefont {Ezawa}\ and\ \citenamefont
  {Iwazaki}(1994)}]{Ezawa1994}%
  \BibitemOpen
  \bibfield  {author} {\bibinfo {author} {\bibfnamefont {Z.}~\bibnamefont
  {Ezawa}}\ and\ \bibinfo {author} {\bibfnamefont {A.}~\bibnamefont
  {Iwazaki}},\ }\href {\doibase 10.1142/S0217979294000877} {\bibfield
  {journal} {\bibinfo  {journal} {International Journal of Modern Physics B}\
  }\textbf {\bibinfo {volume} {08}},\ \bibinfo {pages} {2111} (\bibinfo {year}
  {1994})}\BibitemShut {NoStop}%
\bibitem [{\citenamefont {Moon}\ \emph {et~al.}(1995)\citenamefont {Moon},
  \citenamefont {Mori}, \citenamefont {Yang}, \citenamefont {Girvin},
  \citenamefont {MacDonald}, \citenamefont {Zheng}, \citenamefont {Yoshioka},\
  and\ \citenamefont {Zhang}}]{Moon1995}%
  \BibitemOpen
  \bibfield  {author} {\bibinfo {author} {\bibfnamefont {K.}~\bibnamefont
  {Moon}}, \bibinfo {author} {\bibfnamefont {H.}~\bibnamefont {Mori}}, \bibinfo
  {author} {\bibfnamefont {K.}~\bibnamefont {Yang}}, \bibinfo {author}
  {\bibfnamefont {S.~M.}\ \bibnamefont {Girvin}}, \bibinfo {author}
  {\bibfnamefont {A.~H.}\ \bibnamefont {MacDonald}}, \bibinfo {author}
  {\bibfnamefont {L.}~\bibnamefont {Zheng}}, \bibinfo {author} {\bibfnamefont
  {D.}~\bibnamefont {Yoshioka}}, \ and\ \bibinfo {author} {\bibfnamefont
  {S.-C.}\ \bibnamefont {Zhang}},\ }\href {\doibase 10.1103/PhysRevB.51.5138}
  {\bibfield  {journal} {\bibinfo  {journal} {Phys. Rev. B}\ }\textbf {\bibinfo
  {volume} {51}},\ \bibinfo {pages} {5138} (\bibinfo {year}
  {1995})}\BibitemShut {NoStop}%
\bibitem [{\citenamefont {Yang}\ \emph {et~al.}(1996)\citenamefont {Yang},
  \citenamefont {Moon}, \citenamefont {Belkhir}, \citenamefont {Mori},
  \citenamefont {Girvin}, \citenamefont {MacDonald}, \citenamefont {Zheng},\
  and\ \citenamefont {Yoshioka}}]{Yang1996}%
  \BibitemOpen
  \bibfield  {author} {\bibinfo {author} {\bibfnamefont {K.}~\bibnamefont
  {Yang}}, \bibinfo {author} {\bibfnamefont {K.}~\bibnamefont {Moon}}, \bibinfo
  {author} {\bibfnamefont {L.}~\bibnamefont {Belkhir}}, \bibinfo {author}
  {\bibfnamefont {H.}~\bibnamefont {Mori}}, \bibinfo {author} {\bibfnamefont
  {S.~M.}\ \bibnamefont {Girvin}}, \bibinfo {author} {\bibfnamefont {A.~H.}\
  \bibnamefont {MacDonald}}, \bibinfo {author} {\bibfnamefont {L.}~\bibnamefont
  {Zheng}}, \ and\ \bibinfo {author} {\bibfnamefont {D.}~\bibnamefont
  {Yoshioka}},\ }\href {\doibase 10.1103/PhysRevB.54.11644} {\bibfield
  {journal} {\bibinfo  {journal} {Phys. Rev. B}\ }\textbf {\bibinfo {volume}
  {54}},\ \bibinfo {pages} {11644} (\bibinfo {year} {1996})}\BibitemShut
  {NoStop}%
\bibitem [{\citenamefont {Stern}\ \emph {et~al.}(2001)\citenamefont {Stern},
  \citenamefont {Girvin}, \citenamefont {MacDonald},\ and\ \citenamefont
  {Ma}}]{Stern2001}%
  \BibitemOpen
  \bibfield  {author} {\bibinfo {author} {\bibfnamefont {A.}~\bibnamefont
  {Stern}}, \bibinfo {author} {\bibfnamefont {S.~M.}\ \bibnamefont {Girvin}},
  \bibinfo {author} {\bibfnamefont {A.~H.}\ \bibnamefont {MacDonald}}, \ and\
  \bibinfo {author} {\bibfnamefont {N.}~\bibnamefont {Ma}},\ }\href {\doibase
  10.1103/PhysRevLett.86.1829} {\bibfield  {journal} {\bibinfo  {journal}
  {Phys. Rev. Lett.}\ }\textbf {\bibinfo {volume} {86}},\ \bibinfo {pages}
  {1829} (\bibinfo {year} {2001})}\BibitemShut {NoStop}%
\bibitem [{\citenamefont {Balents}\ and\ \citenamefont
  {Radzihovsky}(2001)}]{Balents2001}%
  \BibitemOpen
  \bibfield  {author} {\bibinfo {author} {\bibfnamefont {L.}~\bibnamefont
  {Balents}}\ and\ \bibinfo {author} {\bibfnamefont {L.}~\bibnamefont
  {Radzihovsky}},\ }\href {\doibase 10.1103/PhysRevLett.86.1825} {\bibfield
  {journal} {\bibinfo  {journal} {Phys. Rev. Lett.}\ }\textbf {\bibinfo
  {volume} {86}},\ \bibinfo {pages} {1825} (\bibinfo {year}
  {2001})}\BibitemShut {NoStop}%
\bibitem [{\citenamefont {Fogler}\ and\ \citenamefont
  {Wilczek}(2001)}]{Fogler2001}%
  \BibitemOpen
  \bibfield  {author} {\bibinfo {author} {\bibfnamefont {M.~M.}\ \bibnamefont
  {Fogler}}\ and\ \bibinfo {author} {\bibfnamefont {F.}~\bibnamefont
  {Wilczek}},\ }\href {\doibase 10.1103/PhysRevLett.86.1833} {\bibfield
  {journal} {\bibinfo  {journal} {Phys. Rev. Lett.}\ }\textbf {\bibinfo
  {volume} {86}},\ \bibinfo {pages} {1833} (\bibinfo {year}
  {2001})}\BibitemShut {NoStop}%
\bibitem [{\citenamefont {Joglekar}\ and\ \citenamefont
  {MacDonald}(2001)}]{Joglekar2001}%
  \BibitemOpen
  \bibfield  {author} {\bibinfo {author} {\bibfnamefont {Y.~N.}\ \bibnamefont
  {Joglekar}}\ and\ \bibinfo {author} {\bibfnamefont {A.~H.}\ \bibnamefont
  {MacDonald}},\ }\href {\doibase 10.1103/PhysRevLett.87.196802} {\bibfield
  {journal} {\bibinfo  {journal} {Phys. Rev. Lett.}\ }\textbf {\bibinfo
  {volume} {87}},\ \bibinfo {pages} {196802} (\bibinfo {year}
  {2001})}\BibitemShut {NoStop}%
\bibitem [{\citenamefont {Yang}(2001)}]{Yang2001}%
  \BibitemOpen
  \bibfield  {author} {\bibinfo {author} {\bibfnamefont {K.}~\bibnamefont
  {Yang}},\ }\href {\doibase 10.1103/PhysRevLett.87.056802} {\bibfield
  {journal} {\bibinfo  {journal} {Phys. Rev. Lett.}\ }\textbf {\bibinfo
  {volume} {87}},\ \bibinfo {pages} {056802} (\bibinfo {year}
  {2001})}\BibitemShut {NoStop}%
\bibitem [{\citenamefont {Demler}\ \emph {et~al.}(2001)\citenamefont {Demler},
  \citenamefont {Nayak},\ and\ \citenamefont {Das~Sarma}}]{Demler2001}%
  \BibitemOpen
  \bibfield  {author} {\bibinfo {author} {\bibfnamefont {E.}~\bibnamefont
  {Demler}}, \bibinfo {author} {\bibfnamefont {C.}~\bibnamefont {Nayak}}, \
  and\ \bibinfo {author} {\bibfnamefont {S.}~\bibnamefont {Das~Sarma}},\ }\href
  {\doibase 10.1103/PhysRevLett.86.1853} {\bibfield  {journal} {\bibinfo
  {journal} {Phys. Rev. Lett.}\ }\textbf {\bibinfo {volume} {86}},\ \bibinfo
  {pages} {1853} (\bibinfo {year} {2001})}\BibitemShut {NoStop}%
\bibitem [{\citenamefont {Stern}\ and\ \citenamefont
  {Halperin}(2002)}]{Stern2002}%
  \BibitemOpen
  \bibfield  {author} {\bibinfo {author} {\bibfnamefont {A.}~\bibnamefont
  {Stern}}\ and\ \bibinfo {author} {\bibfnamefont {B.~I.}\ \bibnamefont
  {Halperin}},\ }\href {\doibase 10.1103/PhysRevLett.88.106801} {\bibfield
  {journal} {\bibinfo  {journal} {Phys. Rev. Lett.}\ }\textbf {\bibinfo
  {volume} {88}},\ \bibinfo {pages} {106801} (\bibinfo {year}
  {2002})}\BibitemShut {NoStop}%
\bibitem [{\citenamefont {Fertig}\ and\ \citenamefont
  {Straley}(2003)}]{Fertig2003}%
  \BibitemOpen
  \bibfield  {author} {\bibinfo {author} {\bibfnamefont {H.~A.}\ \bibnamefont
  {Fertig}}\ and\ \bibinfo {author} {\bibfnamefont {J.~P.}\ \bibnamefont
  {Straley}},\ }\href {\doibase 10.1103/PhysRevLett.91.046806} {\bibfield
  {journal} {\bibinfo  {journal} {Phys. Rev. Lett.}\ }\textbf {\bibinfo
  {volume} {91}},\ \bibinfo {pages} {046806} (\bibinfo {year}
  {2003})}\BibitemShut {NoStop}%
\bibitem [{\citenamefont {Sheng}\ \emph {et~al.}(2003)\citenamefont {Sheng},
  \citenamefont {Balents},\ and\ \citenamefont {Wang}}]{Sheng2003}%
  \BibitemOpen
  \bibfield  {author} {\bibinfo {author} {\bibfnamefont {D.~N.}\ \bibnamefont
  {Sheng}}, \bibinfo {author} {\bibfnamefont {L.}~\bibnamefont {Balents}}, \
  and\ \bibinfo {author} {\bibfnamefont {Z.}~\bibnamefont {Wang}},\ }\href
  {\doibase 10.1103/PhysRevLett.91.116802} {\bibfield  {journal} {\bibinfo
  {journal} {Phys. Rev. Lett.}\ }\textbf {\bibinfo {volume} {91}},\ \bibinfo
  {pages} {116802} (\bibinfo {year} {2003})}\BibitemShut {NoStop}%
\bibitem [{\citenamefont {Fertig}\ and\ \citenamefont
  {Murthy}(2005)}]{Fertig2005}%
  \BibitemOpen
  \bibfield  {author} {\bibinfo {author} {\bibfnamefont {H.~A.}\ \bibnamefont
  {Fertig}}\ and\ \bibinfo {author} {\bibfnamefont {G.}~\bibnamefont
  {Murthy}},\ }\href {\doibase 10.1103/PhysRevLett.95.156802} {\bibfield
  {journal} {\bibinfo  {journal} {Phys. Rev. Lett.}\ }\textbf {\bibinfo
  {volume} {95}},\ \bibinfo {pages} {156802} (\bibinfo {year}
  {2005})}\BibitemShut {NoStop}%
\bibitem [{\citenamefont {Rossi}\ \emph {et~al.}(2005)\citenamefont {Rossi},
  \citenamefont {N\'u\~nez},\ and\ \citenamefont {MacDonald}}]{Rossi2005}%
  \BibitemOpen
  \bibfield  {author} {\bibinfo {author} {\bibfnamefont {E.}~\bibnamefont
  {Rossi}}, \bibinfo {author} {\bibfnamefont {A.~S.}\ \bibnamefont
  {N\'u\~nez}}, \ and\ \bibinfo {author} {\bibfnamefont {A.~H.}\ \bibnamefont
  {MacDonald}},\ }\href {\doibase 10.1103/PhysRevLett.95.266804} {\bibfield
  {journal} {\bibinfo  {journal} {Phys. Rev. Lett.}\ }\textbf {\bibinfo
  {volume} {95}},\ \bibinfo {pages} {266804} (\bibinfo {year}
  {2005})}\BibitemShut {NoStop}%
\bibitem [{\citenamefont {Park}\ and\ \citenamefont
  {Das~Sarma}(2006)}]{Park2006}%
  \BibitemOpen
  \bibfield  {author} {\bibinfo {author} {\bibfnamefont {K.}~\bibnamefont
  {Park}}\ and\ \bibinfo {author} {\bibfnamefont {S.}~\bibnamefont
  {Das~Sarma}},\ }\href {\doibase 10.1103/PhysRevB.74.035338} {\bibfield
  {journal} {\bibinfo  {journal} {Phys. Rev. B}\ }\textbf {\bibinfo {volume}
  {74}},\ \bibinfo {pages} {035338} (\bibinfo {year} {2006})}\BibitemShut
  {NoStop}%
\bibitem [{\citenamefont {Su}\ and\ \citenamefont {MacDonald}(2008)}]{Su2008}%
  \BibitemOpen
  \bibfield  {author} {\bibinfo {author} {\bibfnamefont {J.-J.}\ \bibnamefont
  {Su}}\ and\ \bibinfo {author} {\bibfnamefont {A.}~\bibnamefont {MacDonald}},\
  }\href@noop {} {\bibfield  {journal} {\bibinfo  {journal} {Nature Physics}\
  }\textbf {\bibinfo {volume} {4}},\ \bibinfo {pages} {799} (\bibinfo {year}
  {2008})}\BibitemShut {NoStop}%
\bibitem [{\citenamefont {Eastham}\ \emph {et~al.}(2009)\citenamefont
  {Eastham}, \citenamefont {Cooper},\ and\ \citenamefont {Lee}}]{Eastham2009}%
  \BibitemOpen
  \bibfield  {author} {\bibinfo {author} {\bibfnamefont {P.~R.}\ \bibnamefont
  {Eastham}}, \bibinfo {author} {\bibfnamefont {N.~R.}\ \bibnamefont {Cooper}},
  \ and\ \bibinfo {author} {\bibfnamefont {D.~K.~K.}\ \bibnamefont {Lee}},\
  }\href {\doibase 10.1103/PhysRevB.80.045302} {\bibfield  {journal} {\bibinfo
  {journal} {Phys. Rev. B}\ }\textbf {\bibinfo {volume} {80}},\ \bibinfo
  {pages} {045302} (\bibinfo {year} {2009})}\BibitemShut {NoStop}%
\bibitem [{\citenamefont {Fil}\ and\ \citenamefont
  {Shevchenko}(2009)}]{Fil2009}%
  \BibitemOpen
  \bibfield  {author} {\bibinfo {author} {\bibfnamefont {D.~V.}\ \bibnamefont
  {Fil}}\ and\ \bibinfo {author} {\bibfnamefont {S.~I.}\ \bibnamefont
  {Shevchenko}},\ }\href {http://stacks.iop.org/0953-8984/21/i=21/a=215701}
  {\bibfield  {journal} {\bibinfo  {journal} {Journal of Physics: Condensed
  Matter}\ }\textbf {\bibinfo {volume} {21}},\ \bibinfo {pages} {215701}
  (\bibinfo {year} {2009})}\BibitemShut {NoStop}%
\bibitem [{\citenamefont {Dolcini}\ \emph {et~al.}(2010)\citenamefont
  {Dolcini}, \citenamefont {Rainis}, \citenamefont {Taddei}, \citenamefont
  {Polini}, \citenamefont {Fazio},\ and\ \citenamefont
  {MacDonald}}]{Dolcini2010}%
  \BibitemOpen
  \bibfield  {author} {\bibinfo {author} {\bibfnamefont {F.}~\bibnamefont
  {Dolcini}}, \bibinfo {author} {\bibfnamefont {D.}~\bibnamefont {Rainis}},
  \bibinfo {author} {\bibfnamefont {F.}~\bibnamefont {Taddei}}, \bibinfo
  {author} {\bibfnamefont {M.}~\bibnamefont {Polini}}, \bibinfo {author}
  {\bibfnamefont {R.}~\bibnamefont {Fazio}}, \ and\ \bibinfo {author}
  {\bibfnamefont {A.~H.}\ \bibnamefont {MacDonald}},\ }\href {\doibase
  10.1103/PhysRevLett.104.027004} {\bibfield  {journal} {\bibinfo  {journal}
  {Phys. Rev. Lett.}\ }\textbf {\bibinfo {volume} {104}},\ \bibinfo {pages}
  {027004} (\bibinfo {year} {2010})}\BibitemShut {NoStop}%
\bibitem [{\citenamefont {Eastham}\ \emph {et~al.}(2010)\citenamefont
  {Eastham}, \citenamefont {Cooper},\ and\ \citenamefont {Lee}}]{Eastham2010}%
  \BibitemOpen
  \bibfield  {author} {\bibinfo {author} {\bibfnamefont {P.~R.}\ \bibnamefont
  {Eastham}}, \bibinfo {author} {\bibfnamefont {N.~R.}\ \bibnamefont {Cooper}},
  \ and\ \bibinfo {author} {\bibfnamefont {D.~K.~K.}\ \bibnamefont {Lee}},\
  }\href {\doibase 10.1103/PhysRevLett.105.236805} {\bibfield  {journal}
  {\bibinfo  {journal} {Phys. Rev. Lett.}\ }\textbf {\bibinfo {volume} {105}},\
  \bibinfo {pages} {236805} (\bibinfo {year} {2010})}\BibitemShut {NoStop}%
\bibitem [{\citenamefont {Su}\ and\ \citenamefont {MacDonald}(2010)}]{Su2010}%
  \BibitemOpen
  \bibfield  {author} {\bibinfo {author} {\bibfnamefont {J.-J.}\ \bibnamefont
  {Su}}\ and\ \bibinfo {author} {\bibfnamefont {A.~H.}\ \bibnamefont
  {MacDonald}},\ }\href {\doibase 10.1103/PhysRevB.81.184523} {\bibfield
  {journal} {\bibinfo  {journal} {Phys. Rev. B}\ }\textbf {\bibinfo {volume}
  {81}},\ \bibinfo {pages} {184523} (\bibinfo {year} {2010})}\BibitemShut
  {NoStop}%
\bibitem [{\citenamefont {Hyart}\ and\ \citenamefont
  {Rosenow}(2011)}]{Hyart2011}%
  \BibitemOpen
  \bibfield  {author} {\bibinfo {author} {\bibfnamefont {T.}~\bibnamefont
  {Hyart}}\ and\ \bibinfo {author} {\bibfnamefont {B.}~\bibnamefont
  {Rosenow}},\ }\href {\doibase 10.1103/PhysRevB.83.155315} {\bibfield
  {journal} {\bibinfo  {journal} {Phys. Rev. B}\ }\textbf {\bibinfo {volume}
  {83}},\ \bibinfo {pages} {155315} (\bibinfo {year} {2011})}\BibitemShut
  {NoStop}%
\bibitem [{\citenamefont {Pesin}\ and\ \citenamefont
  {MacDonald}(2011)}]{Pesin2011}%
  \BibitemOpen
  \bibfield  {author} {\bibinfo {author} {\bibfnamefont {D.~A.}\ \bibnamefont
  {Pesin}}\ and\ \bibinfo {author} {\bibfnamefont {A.~H.}\ \bibnamefont
  {MacDonald}},\ }\href {\doibase 10.1103/PhysRevB.84.075308} {\bibfield
  {journal} {\bibinfo  {journal} {Phys. Rev. B}\ }\textbf {\bibinfo {volume}
  {84}},\ \bibinfo {pages} {075308} (\bibinfo {year} {2011})}\BibitemShut
  {NoStop}%
\bibitem [{\citenamefont {Hyart}\ and\ \citenamefont
  {Rosenow}(2013)}]{Hyart2013}%
  \BibitemOpen
  \bibfield  {author} {\bibinfo {author} {\bibfnamefont {T.}~\bibnamefont
  {Hyart}}\ and\ \bibinfo {author} {\bibfnamefont {B.}~\bibnamefont
  {Rosenow}},\ }\href {\doibase 10.1103/PhysRevLett.110.076806} {\bibfield
  {journal} {\bibinfo  {journal} {Phys. Rev. Lett.}\ }\textbf {\bibinfo
  {volume} {110}},\ \bibinfo {pages} {076806} (\bibinfo {year}
  {2013})}\BibitemShut {NoStop}%
\bibitem [{\citenamefont {Hama}\ \emph {et~al.}(2013)\citenamefont {Hama},
  \citenamefont {Tsitsishvili},\ and\ \citenamefont {Ezawa}}]{Hama2013}%
  \BibitemOpen
  \bibfield  {author} {\bibinfo {author} {\bibfnamefont {Y.}~\bibnamefont
  {Hama}}, \bibinfo {author} {\bibfnamefont {G.}~\bibnamefont {Tsitsishvili}},
  \ and\ \bibinfo {author} {\bibfnamefont {Z.~F.}\ \bibnamefont {Ezawa}},\
  }\href {\doibase 10.1093/ptep/ptt025} {\bibfield  {journal} {\bibinfo
  {journal} {Progress of Theoretical and Experimental Physics}\ }\textbf
  {\bibinfo {volume} {2013}} (\bibinfo {year} {2013}),\
  10.1093/ptep/ptt025}\BibitemShut {NoStop}%
\bibitem [{\citenamefont {Spielman}\ \emph {et~al.}(2001)\citenamefont
  {Spielman}, \citenamefont {Eisenstein}, \citenamefont {Pfeiffer},\ and\
  \citenamefont {West}}]{Spielman2001}%
  \BibitemOpen
  \bibfield  {author} {\bibinfo {author} {\bibfnamefont {I.~B.}\ \bibnamefont
  {Spielman}}, \bibinfo {author} {\bibfnamefont {J.~P.}\ \bibnamefont
  {Eisenstein}}, \bibinfo {author} {\bibfnamefont {L.~N.}\ \bibnamefont
  {Pfeiffer}}, \ and\ \bibinfo {author} {\bibfnamefont {K.~W.}\ \bibnamefont
  {West}},\ }\href {\doibase 10.1103/PhysRevLett.87.036803} {\bibfield
  {journal} {\bibinfo  {journal} {Phys. Rev. Lett.}\ }\textbf {\bibinfo
  {volume} {87}},\ \bibinfo {pages} {036803} (\bibinfo {year}
  {2001})}\BibitemShut {NoStop}%
\bibitem [{\citenamefont {Spielman}(2004)}]{Spielman2004}%
  \BibitemOpen
  \bibfield  {author} {\bibinfo {author} {\bibfnamefont {I.~B.}\ \bibnamefont
  {Spielman}},\ }\emph {\bibinfo {title} {Evidence for the Josephson Effect in
  Quantum Hall Bilayers}},\ \href@noop {} {Ph.D. thesis},\ \bibinfo  {school}
  {Caltech} (\bibinfo {year} {2004})\BibitemShut {NoStop}%
\bibitem [{\citenamefont {Tiemann}\ \emph {et~al.}(2008)\citenamefont
  {Tiemann}, \citenamefont {Dietsche}, \citenamefont {Hauser},\ and\
  \citenamefont {von Klitzing}}]{Tiemann2008}%
  \BibitemOpen
  \bibfield  {author} {\bibinfo {author} {\bibfnamefont {L.}~\bibnamefont
  {Tiemann}}, \bibinfo {author} {\bibfnamefont {W.}~\bibnamefont {Dietsche}},
  \bibinfo {author} {\bibfnamefont {M.}~\bibnamefont {Hauser}}, \ and\ \bibinfo
  {author} {\bibfnamefont {K.}~\bibnamefont {von Klitzing}},\ }\href@noop {}
  {\bibfield  {journal} {\bibinfo  {journal} {New Journal of Physics}\ }\textbf
  {\bibinfo {volume} {10}},\ \bibinfo {pages} {045018} (\bibinfo {year}
  {2008})}\BibitemShut {NoStop}%
\bibitem [{\citenamefont {Tiemann}\ \emph {et~al.}(2009)\citenamefont
  {Tiemann}, \citenamefont {Yoon}, \citenamefont {Dietsche}, \citenamefont {von
  Klitzing},\ and\ \citenamefont {Wegscheider}}]{Tiemann2009}%
  \BibitemOpen
  \bibfield  {author} {\bibinfo {author} {\bibfnamefont {L.}~\bibnamefont
  {Tiemann}}, \bibinfo {author} {\bibfnamefont {Y.}~\bibnamefont {Yoon}},
  \bibinfo {author} {\bibfnamefont {W.}~\bibnamefont {Dietsche}}, \bibinfo
  {author} {\bibfnamefont {K.}~\bibnamefont {von Klitzing}}, \ and\ \bibinfo
  {author} {\bibfnamefont {W.}~\bibnamefont {Wegscheider}},\ }\href {\doibase
  10.1103/PhysRevB.80.165120} {\bibfield  {journal} {\bibinfo  {journal} {Phys.
  Rev. B}\ }\textbf {\bibinfo {volume} {80}},\ \bibinfo {pages} {165120}
  (\bibinfo {year} {2009})}\BibitemShut {NoStop}%
\bibitem [{\citenamefont {Yoon}\ \emph {et~al.}(2010)\citenamefont {Yoon},
  \citenamefont {Tiemann}, \citenamefont {Schmult}, \citenamefont {Dietsche},
  \citenamefont {von Klitzing},\ and\ \citenamefont {Wegscheider}}]{Yoon2010}%
  \BibitemOpen
  \bibfield  {author} {\bibinfo {author} {\bibfnamefont {Y.}~\bibnamefont
  {Yoon}}, \bibinfo {author} {\bibfnamefont {L.}~\bibnamefont {Tiemann}},
  \bibinfo {author} {\bibfnamefont {S.}~\bibnamefont {Schmult}}, \bibinfo
  {author} {\bibfnamefont {W.}~\bibnamefont {Dietsche}}, \bibinfo {author}
  {\bibfnamefont {K.}~\bibnamefont {von Klitzing}}, \ and\ \bibinfo {author}
  {\bibfnamefont {W.}~\bibnamefont {Wegscheider}},\ }\href {\doibase
  10.1103/PhysRevLett.104.116802} {\bibfield  {journal} {\bibinfo  {journal}
  {Phys. Rev. Lett.}\ }\textbf {\bibinfo {volume} {104}},\ \bibinfo {pages}
  {116802} (\bibinfo {year} {2010})}\BibitemShut {NoStop}%
\bibitem [{\citenamefont {Huang}\ \emph {et~al.}(2012)\citenamefont {Huang},
  \citenamefont {Dietsche}, \citenamefont {Hauser},\ and\ \citenamefont {von
  Klitzing}}]{Huang2012}%
  \BibitemOpen
  \bibfield  {author} {\bibinfo {author} {\bibfnamefont {X.}~\bibnamefont
  {Huang}}, \bibinfo {author} {\bibfnamefont {W.}~\bibnamefont {Dietsche}},
  \bibinfo {author} {\bibfnamefont {M.}~\bibnamefont {Hauser}}, \ and\ \bibinfo
  {author} {\bibfnamefont {K.}~\bibnamefont {von Klitzing}},\ }\href {\doibase
  10.1103/PhysRevLett.109.156802} {\bibfield  {journal} {\bibinfo  {journal}
  {Phys. Rev. Lett.}\ }\textbf {\bibinfo {volume} {109}},\ \bibinfo {pages}
  {156802} (\bibinfo {year} {2012})}\BibitemShut {NoStop}%
\bibitem [{\citenamefont {Nandi}\ \emph {et~al.}(2013)\citenamefont {Nandi},
  \citenamefont {Khaire}, \citenamefont {Finck}, \citenamefont {Eisenstein},
  \citenamefont {Pfeiffer},\ and\ \citenamefont {West}}]{Nandi2013}%
  \BibitemOpen
  \bibfield  {author} {\bibinfo {author} {\bibfnamefont {D.}~\bibnamefont
  {Nandi}}, \bibinfo {author} {\bibfnamefont {T.}~\bibnamefont {Khaire}},
  \bibinfo {author} {\bibfnamefont {A.~D.~K.}\ \bibnamefont {Finck}}, \bibinfo
  {author} {\bibfnamefont {J.~P.}\ \bibnamefont {Eisenstein}}, \bibinfo
  {author} {\bibfnamefont {L.~N.}\ \bibnamefont {Pfeiffer}}, \ and\ \bibinfo
  {author} {\bibfnamefont {K.~W.}\ \bibnamefont {West}},\ }\href {\doibase
  10.1103/PhysRevB.88.165308} {\bibfield  {journal} {\bibinfo  {journal} {Phys.
  Rev. B}\ }\textbf {\bibinfo {volume} {88}},\ \bibinfo {pages} {165308}
  (\bibinfo {year} {2013})}\BibitemShut {NoStop}%
\bibitem [{\citenamefont {Zhang}\ \emph {et~al.}(2014)\citenamefont {Zhang},
  \citenamefont {Huang}, \citenamefont {Dietsche}, \citenamefont {Hauser},\
  and\ \citenamefont {von Klitzing}}]{Zhang2014}%
  \BibitemOpen
  \bibfield  {author} {\bibinfo {author} {\bibfnamefont {D.}~\bibnamefont
  {Zhang}}, \bibinfo {author} {\bibfnamefont {X.}~\bibnamefont {Huang}},
  \bibinfo {author} {\bibfnamefont {W.}~\bibnamefont {Dietsche}}, \bibinfo
  {author} {\bibfnamefont {M.}~\bibnamefont {Hauser}}, \ and\ \bibinfo {author}
  {\bibfnamefont {K.}~\bibnamefont {von Klitzing}},\ }\href {\doibase
  10.1103/PhysRevB.90.085436} {\bibfield  {journal} {\bibinfo  {journal} {Phys.
  Rev. B}\ }\textbf {\bibinfo {volume} {90}},\ \bibinfo {pages} {085436}
  (\bibinfo {year} {2014})}\BibitemShut {NoStop}%
\bibitem [{Note1()}]{Note1}%
  \BibitemOpen
  \bibinfo {note} {This is a direct consequence of the fact that $\phi $ and
  $M_z$ are canonically conjugated variables~\cite {Wen1993}: the
  Euler-Lagrange equation for $M_z$ implies $\hbar \protect \mathaccentV
  {dot}05F{\phi }=\delta \protect \mathcal {E}/\delta M_z$, and the chemical
  potential difference between the layers is the energy cost of moving one
  electron from the bottom to the top layer, for which $\DOTSI \intop \ilimits@
  d^2 r \ \delta M_z=1$.}\BibitemShut {Stop}%
\bibitem [{\citenamefont {Gilbert}(2004)}]{Gilbert2004}%
  \BibitemOpen
  \bibfield  {author} {\bibinfo {author} {\bibfnamefont {T.~L.}\ \bibnamefont
  {Gilbert}},\ }\href@noop {} {\bibfield  {journal} {\bibinfo  {journal}
  {Magnetics, IEEE Transactions on}\ }\textbf {\bibinfo {volume} {40}},\
  \bibinfo {pages} {3443} (\bibinfo {year} {2004})}\BibitemShut {NoStop}%
\bibitem [{\citenamefont {Tinkham}(2012)}]{Tinkham2012}%
  \BibitemOpen
  \bibfield  {author} {\bibinfo {author} {\bibfnamefont {M.}~\bibnamefont
  {Tinkham}},\ }\href@noop {} {\emph {\bibinfo {title} {Introduction to
  superconductivity}}}\ (\bibinfo  {publisher} {Courier Dover Publications},\
  \bibinfo {year} {2012})\BibitemShut {NoStop}%
\bibitem [{\citenamefont {Prange}\ and\ \citenamefont
  {Girvin}(1987)}]{Prange1987}%
  \BibitemOpen
  \bibinfo {editor} {\bibfnamefont {R.}~\bibnamefont {Prange}}\ and\ \bibinfo
  {editor} {\bibfnamefont {S.}~\bibnamefont {Girvin}},\ eds.,\ \href@noop {}
  {\emph {\bibinfo {title} {The quantum Hall effect}}},\ Vol.~\bibinfo {volume}
  {1}\ (\bibinfo  {publisher} {Springer-Verlag New York},\ \bibinfo {year}
  {1987})\ Chap.~\bibinfo {chapter} {8}\BibitemShut {NoStop}%
\bibitem [{\citenamefont {Chen}\ \emph {et~al.}(2014)\citenamefont {Chen},
  \citenamefont {Kent}, \citenamefont {MacDonald},\ and\ \citenamefont
  {Sodemann}}]{Chen2014}%
  \BibitemOpen
  \bibfield  {author} {\bibinfo {author} {\bibfnamefont {H.}~\bibnamefont
  {Chen}}, \bibinfo {author} {\bibfnamefont {A.~D.}\ \bibnamefont {Kent}},
  \bibinfo {author} {\bibfnamefont {A.~H.}\ \bibnamefont {MacDonald}}, \ and\
  \bibinfo {author} {\bibfnamefont {I.}~\bibnamefont {Sodemann}},\ }\href@noop
  {} {\bibfield  {journal} {\bibinfo  {journal} {arXiv preprint
  arXiv:1408.1100}\ } (\bibinfo {year} {2014})}\BibitemShut {NoStop}%
\bibitem [{\citenamefont {Tserkovnyak}\ \emph {et~al.}(2002)\citenamefont
  {Tserkovnyak}, \citenamefont {Brataas},\ and\ \citenamefont
  {Bauer}}]{Tserkovnyak2002}%
  \BibitemOpen
  \bibfield  {author} {\bibinfo {author} {\bibfnamefont {Y.}~\bibnamefont
  {Tserkovnyak}}, \bibinfo {author} {\bibfnamefont {A.}~\bibnamefont
  {Brataas}}, \ and\ \bibinfo {author} {\bibfnamefont {G.~E.~W.}\ \bibnamefont
  {Bauer}},\ }\href {\doibase 10.1103/PhysRevLett.88.117601} {\bibfield
  {journal} {\bibinfo  {journal} {Phys. Rev. Lett.}\ }\textbf {\bibinfo
  {volume} {88}},\ \bibinfo {pages} {117601} (\bibinfo {year}
  {2002})}\BibitemShut {NoStop}%
\bibitem [{\citenamefont {Tserkovnyak}\ \emph {et~al.}(2005)\citenamefont
  {Tserkovnyak}, \citenamefont {Brataas}, \citenamefont {Bauer},\ and\
  \citenamefont {Halperin}}]{Tserkovnyak2005}%
  \BibitemOpen
  \bibfield  {author} {\bibinfo {author} {\bibfnamefont {Y.}~\bibnamefont
  {Tserkovnyak}}, \bibinfo {author} {\bibfnamefont {A.}~\bibnamefont
  {Brataas}}, \bibinfo {author} {\bibfnamefont {G.~E.~W.}\ \bibnamefont
  {Bauer}}, \ and\ \bibinfo {author} {\bibfnamefont {B.~I.}\ \bibnamefont
  {Halperin}},\ }\href {\doibase 10.1103/RevModPhys.77.1375} {\bibfield
  {journal} {\bibinfo  {journal} {Rev. Mod. Phys.}\ }\textbf {\bibinfo {volume}
  {77}},\ \bibinfo {pages} {1375} (\bibinfo {year} {2005})}\BibitemShut
  {NoStop}%
\bibitem [{\citenamefont {Finck}\ \emph {et~al.}(2008)\citenamefont {Finck},
  \citenamefont {Champagne}, \citenamefont {Eisenstein}, \citenamefont
  {Pfeiffer},\ and\ \citenamefont {West}}]{Finck2008}%
  \BibitemOpen
  \bibfield  {author} {\bibinfo {author} {\bibfnamefont {A.~D.~K.}\
  \bibnamefont {Finck}}, \bibinfo {author} {\bibfnamefont {A.~R.}\ \bibnamefont
  {Champagne}}, \bibinfo {author} {\bibfnamefont {J.~P.}\ \bibnamefont
  {Eisenstein}}, \bibinfo {author} {\bibfnamefont {L.~N.}\ \bibnamefont
  {Pfeiffer}}, \ and\ \bibinfo {author} {\bibfnamefont {K.~W.}\ \bibnamefont
  {West}},\ }\href {\doibase 10.1103/PhysRevB.78.075302} {\bibfield  {journal}
  {\bibinfo  {journal} {Phys. Rev. B}\ }\textbf {\bibinfo {volume} {78}},\
  \bibinfo {pages} {075302} (\bibinfo {year} {2008})}\BibitemShut {NoStop}%
\bibitem [{Note2()}]{Note2}%
  \BibitemOpen
  \bibinfo {note} {The system I-V curve has a region of negative differential
  resistivity at voltages immediately above the cusp which leads to circuit
  instabilities that make measurements extremely close to the maximum in the
  I-V curve difficult. Ref.~\protect \rev@citealp {Nandi2013} addressed this
  experimental challenge and provided a thorough characterization of the
  behavior of the I-V curve immediately above the critical
  current.}\BibitemShut {Stop}%
\bibitem [{Note3()}]{Note3}%
  \BibitemOpen
  \bibinfo {note} {The minus sign arises because the positive loop resistance
  sign convention has been redefined to be be positive when the number current
  is injected into the bottom layer to match the convention of Ref.~\cite
  {Yoon2010}.}\BibitemShut {Stop}%
\bibitem [{Note4()}]{Note4}%
  \BibitemOpen
  \bibinfo {note} {In this sense the very observation of the independence of
  the critical current on the contact position along the rim of the Corbino
  annulus is already an indication that $\lambda $ is longer than the system
  size~\cite {Nandi2013}.}\BibitemShut {Stop}%
\bibitem [{\citenamefont {Grimes}\ and\ \citenamefont
  {Shapiro}(1968)}]{Grimes1968}%
  \BibitemOpen
  \bibfield  {author} {\bibinfo {author} {\bibfnamefont {C.~C.}\ \bibnamefont
  {Grimes}}\ and\ \bibinfo {author} {\bibfnamefont {S.}~\bibnamefont
  {Shapiro}},\ }\href {\doibase 10.1103/PhysRev.169.397} {\bibfield  {journal}
  {\bibinfo  {journal} {Phys. Rev.}\ }\textbf {\bibinfo {volume} {169}},\
  \bibinfo {pages} {397} (\bibinfo {year} {1968})}\BibitemShut {NoStop}%
\bibitem [{\citenamefont {Renne}\ and\ \citenamefont
  {Polder}(1974)}]{Renne1974}%
  \BibitemOpen
  \bibfield  {author} {\bibinfo {author} {\bibfnamefont {M.}~\bibnamefont
  {Renne}}\ and\ \bibinfo {author} {\bibfnamefont {D.}~\bibnamefont {Polder}},\
  }\href@noop {} {\bibfield  {journal} {\bibinfo  {journal} {Revue de physique
  appliqu{\'e}e}\ }\textbf {\bibinfo {volume} {9}},\ \bibinfo {pages} {25}
  (\bibinfo {year} {1974})}\BibitemShut {NoStop}%
\bibitem [{\citenamefont {Sondhi}\ \emph {et~al.}(1993)\citenamefont {Sondhi},
  \citenamefont {Karlhede}, \citenamefont {Kivelson},\ and\ \citenamefont
  {Rezayi}}]{Sondhi1993}%
  \BibitemOpen
  \bibfield  {author} {\bibinfo {author} {\bibfnamefont {S.~L.}\ \bibnamefont
  {Sondhi}}, \bibinfo {author} {\bibfnamefont {A.}~\bibnamefont {Karlhede}},
  \bibinfo {author} {\bibfnamefont {S.~A.}\ \bibnamefont {Kivelson}}, \ and\
  \bibinfo {author} {\bibfnamefont {E.~H.}\ \bibnamefont {Rezayi}},\ }\href
  {\doibase 10.1103/PhysRevB.47.16419} {\bibfield  {journal} {\bibinfo
  {journal} {Phys. Rev. B}\ }\textbf {\bibinfo {volume} {47}},\ \bibinfo
  {pages} {16419} (\bibinfo {year} {1993})}\BibitemShut {NoStop}%
\end{thebibliography}
\end{document}